\documentclass[aps,prc,twocolumn,floatfix,nofootinbib,showpacs,superscriptaddress]{revtex4-1}

\usepackage{amsmath,amsfonts,amssymb,bm}
\usepackage{graphicx}
\usepackage{multirow}
\usepackage{color}
\usepackage{pzccal}
\usepackage{soul}
\definecolor{purple}{rgb}{0.5,0,0.5}
\definecolor{blue}{rgb}{0.0,0,0.9}

\begin{document}

\title{Features and flaws of a contact interaction treatment of the kaon}

\author{Chen Chen}
\affiliation{Institute for Theoretical Physics and Department of Modern Physics,
University of Science and Technology of China, Hefei 230026, P.\ R.\ China}
\affiliation{Physics Division, Argonne National Laboratory, Argonne, Illinois 60439, USA}
\affiliation{Department of Physics, Illinois Institute of Technology, Chicago, Illinois 60616-3793, USA}

\author{Lei Chang}
\affiliation{Institut f\"ur Kernphysik, Forschungszentrum J\"ulich, D-52425 J\"ulich, Germany}

\author{\mbox{Craig D.~Roberts}}
\affiliation{Physics Division, Argonne National Laboratory, Argonne, Illinois 60439, USA}
\affiliation{Department of Physics, Illinois Institute of Technology, Chicago, Illinois 60616-3793, USA}

\author{Sebastian M.~Schmidt}
\affiliation{Institute for Advanced Simulation, Forschungszentrum J\"ulich and JARA, D-52425 J\"ulich, Germany}

\author{Shaolong Wan}
\affiliation{Institute for Theoretical Physics and Department of Modern Physics,
University of Science and Technology of China, Hefei 230026, P.\ R.\ China}

\author{David J.~Wilson}
\affiliation{Department of Physics, Old Dominion University, Norfolk, VA 23529, USA}

\date{4 December 2012}

\begin{abstract}
Elastic and semileptonic transition form factors for the kaon and pion are calculated using the leading-order in a global-symmetry-preserving truncation of the Dyson-Schwinger equations and a momentum-independent form for the associated kernels in the gap and Bethe-Salpeter equations.
The computed form factors are compared both with those obtained using the same truncation but an interaction that preserves the one-loop renormalisation-group behaviour of QCD and with data.  The comparisons show that: in connection with observables revealed by probes with $|Q^2|\lesssim M^2$, where $M\approx 0.4\,$GeV is an infrared value of the dressed-quark mass, results obtained using a symmetry-preserving regularisation of the contact-interaction are not realistically distinguishable from those produced by more sophisticated kernels; and available data on kaon form factors do not extend into the domain whereupon one could distinguish between the interactions.
The situation is different if one includes the domain $Q^2>M^2$.  Thereupon, a fully consistent treatment of the contact interaction produces form factors that are typically harder than those obtained with QCD renormalisation-group-improved kernels.
Amongst other things also described are a Ward identity for the inhomogeneous scalar vertex, similarity between the charge distribution of a dressed-$u$-quark in the $K^+$ and that of the dressed-$u$-quark in the $\pi^+$, and reflections upon the point whereat one might begin to see perturbative behaviour in the pion form factor.
Interpolations of the form factors are provided, which should assist in working to chart the interaction between light-quarks by explicating the impact on hadron properties of differing assumptions about the behaviour of the Bethe-Salpeter kernel.
\end{abstract}


\maketitle

\section{Introduction}
The concept and definition of strangeness emerged over a roughly twenty-year period, following discovery of the (neutral) kaon \cite{Rochester:1947mi}.  Like the charged pions, all kaons are stable against strong and electromagnetic decays.  Hence, to the first observers they appeared to survive a strangely long time.  As a strong interaction bound state, whose decay is mediated only by the weak interaction (thus, the relatively long lifetime), kaons have been instrumental in establishing the foundation and properties of the Standard Model; most notably, perhaps, the notions of CP violation and weak-interaction mixing between quark flavours.  Simple phenomena involving kaons continue to provide a valuable means by which to make precision tests of the Standard Model \cite{Sciascia:2008fr}.

The association of strangeness with a distinct quark flavour was complete with the advent of the constituent quark model \cite{GellMann:1964nj,Zweig:1981pd}.  As a scheme for systematically classifying hadrons according to their quark content, the quark model is adequate for pions and kaons.  However, it early became clear \cite{GellMann:1968rz,Weinberg:1966kf} that quantum mechanical models cannot veraciously describe the masses and interactions of pions and kaons because they are (pseudo-) Goldstone bosons associated with dynamical chiral symmetry breaking (DCSB).

Owing to the existence of nonzero current-quark masses \cite{Leutwyler:2009jg}, pions and kaons are not consummate Goldstone modes.  Comparison between their properties can expose the differences in magnitudes between both the current- and constituent-like masses of $u,d$-quarks and the $s$-quark; and the wider impact of these differences.  This window on $SU(3)$-flavour symmetry breaking therefore provides direct access to both explicit and dynamical effects in a wide variety of domains.  For example: the mass formulae for pseudo-Goldstone bosons involve both current-quark masses and order parameters for DCSB \cite{Maris:1997tm,Brodsky:2012ku}; and the ratio of kaon and pion valence-$u$-quark distribution functions provides access to a renormalisation scale invariant ratio of DCSB order parameters \cite{Holt:2010vj,Nguyen:2011jy}.


Herein we analyse kaon and pion elastic and semileptonic transition form factors with the framework of QCD's Dyson-Schwinger equations \cite{Roberts:2000aa,Chang:2011vu,Bashir:2012fs}.  This study is an integral part of a larger programme, aimed at charting the interaction between light-quarks by explicating the impact of differing assumptions about the behaviour of the Bethe-Salpeter kernel upon the spectrum of hadrons, and also upon their elastic and transition form factors on a large domain of momentum transfer.
Material progress has been made in connection with $u,d$-quark systems \cite{GutierrezGuerrero:2010md,Roberts:2010rn,Roberts:2011cf,Roberts:2011wy,Wilson:2011aa}; and a spectrum of mesons and baryons with one or more $s$-quarks was recently computed \cite{Chen:2012qr}.  In order to extend the latter to predictions for elastic and transition form factors of baryons containing $s$-quarks, it is necessary to compute such form factors for mesons containing $s$-quarks.

This need is readily explained \cite{Roberts:2011wy}.  Many properties of baryons are successfully described via a Poincar\'e covariant Faddeev equation that expresses the presence of nonpointlike diquark correlations \cite{Cahill:1988dx}, evidence for the existence of which is accumulating \cite{Cloet:2008re,Cloet:2011qu,Cates:2011pz,Wilson:2011aa}.   There is a fundamental mathematical similarity between diquark correlations and uniquely identified meson analogues \cite{Cahill:1987qr}.  Hence, computing the properties of these analogues is principally equivalent to producing results for the diquark correlations \cite{Roberts:2011wy}; and form factors for diquark correlations are a necessary piece in the computation of baryon form factors \cite{Wilson:2011aa}.

In addition, as indicated above, these meson form factors are interesting in their own right.  The pion elastic form factor is much studied theoretically and, on a modest domain, constrained well experimentally \cite{Volmer:2000ek,Huber:2008id,Holt:2012gg}.  However, there are fewer single-framework analyses of the pion transition and kaon form factors in which quark-gluon dynamics is discernible (e.g., Refs.\,\cite{Burden:1995ve,Kalinovsky:1996ii,Choi:1998jd,Maris:2000sk,Ji:2000fy,%
Ji:2001pj,Colangelo:2010et}), and the data is typically old and imprecise \cite{Molzon:1978py,Dally:1980dj,Amendolia:1986ui,Beringer:1900zz}.  We thus focus primarily on the kaon elastic and transition ($K_{\ell 3}$) form factors, and the pion transition ($\pi_{e3}$).  (N.B.\, Except in connection with the pion transition form factor, we assume isospin symmetry.)

As a Poincar\'e-covariant framework, capable of simultaneously implementing light-quark confinement and expressing DCSB, and admitting a symmetry-preserving truncation scheme, the DSEs are an excellent tool for analysis of these form factors involving the pion and kaon pseudo-Goldstone bosons.  Our study will exploit a symmetry-preserving treatment of a vector$\times$vector contact interaction.  In contrasting the behaviour produced by such an interaction with that obtained using a momentum-dependent interaction, which preserves the one-loop renormalisation group behaviour of QCD, we will achieve comparisons that expose those observables which are most sensitive to the infrared evolution of the strong interaction's running coupling and masses or might become so in future with additional experimental effort.  Moreover, from careful interpretation of the contact-interaction results, one can draw additional valuable insights.
To express this differently, we take a global perspective centred on the strong interaction and make no effort to fine tune interaction parameters.  Plainly, then, we do not intend that this study should make a material contribution to precision tests of the Standard model.  Rather, we seek to identify those aspects of kaon and pion physics that can serve to discriminate between conjectures about strong interaction dynamics.

In Sec.\,\ref{sec:background} we introduce the matrix elements that must be computed, present the associated kinematics, provide some background material on the nature of $SU(3)$-flavour symmetry breaking, and present the formulae we use to calculate the matrix elements.  We also detail the impact of symmetries and dynamics on the vector part of the dressed-quark--$W$-boson vertex: an informed understanding of this vertex is crucial in any analysis of hadron form factors.
This section is complemented by three appendices.  They detail our symmetry-preserving treatment of the vector$\times$vector contact interaction, including some information about the gap and Bethe-Salpeter equations and current conservation; and list a formula for the kaon elastic form factor, which is readily mapped into an expression for the analogous pion form factor.
In Secs.\,\ref{sec:resultselastic} and \ref{sec:resultstransition} we present our results along with a detailed comparative analysis in the context of experiment and kindred theoretical studies.
We recapitulate and conclude in Sec.\,\ref{sec:epilogue}.

\section{Form Factors}
\label{sec:background}
\subsection{Definitions}
We are interested in the following matrix elements
\begin{eqnarray}
\nonumber
M_\mu^K(P,Q) & = & \langle K^+(p) | \sum_{f=u,\bar s} e_f\bar q_f i\gamma_\mu q_f |K^+(k)\rangle\\
& = & 2 P_\mu F_K(Q^2)\,, \label{Kelastic}\\
\nonumber
M_\mu^\pi(P,Q) & = & \langle \pi^+(p) | \sum_{f=u,\bar d} e_f\bar q_f i\gamma_\mu q_f |\pi^+(k)\rangle \\
& = & 2 P_\mu F_\pi(Q^2)\,, \label{pionFF}\\
\nonumber
M_\mu^{K_{\ell 3}}(P,Q) &=& \langle \pi^0(p) | \bar s i\gamma_\mu u |K^+(k)\rangle\\
& = & \frac{1}{\surd 2} \big[ f_+^K(Q^2) 2 P_\mu - f_-^K(Q^2)Q_\mu \big]\,,
\label{Kl3FF}\\
\nonumber
M_\mu^{\pi_{e3}}(P,Q) &=& \langle \pi^0(p) | \bar d i\gamma_\mu u |\pi^+(k)\rangle\\
& = & \frac{1}{\surd 2} \big[ f_+^\pi(Q^2) 2 P_\mu - f_-^\pi(Q^2)Q_\mu \big]\,,
\end{eqnarray}
where: $e_u=2/3$, $e_{\bar d}=1/3=e_{\bar s}$;
$2P=k+p$, $Q=p-k$, with $k^2 = -m_K^2,-m_\pi^2$ and $p^2=-m_K^2,-m_\pi^2$, depending on the initial and final state; and the squared-momentum-transfer is $t=-Q^2$.\footnote{We use a Euclidean metric:  $\{\gamma_\mu,\gamma_\nu\} = 2\delta_{\mu\nu}$; $\gamma_\mu^\dagger = \gamma_\mu$; $\gamma_5= \gamma_4\gamma_1\gamma_2\gamma_3$, tr$[\gamma_5\gamma_\mu\gamma_\nu\gamma_\rho\gamma_\sigma]=-4 \epsilon_{\mu\nu\rho\sigma}$; $\sigma_{\mu\nu}=(i/2)[\gamma_\mu,\gamma_\nu]$; $a \cdot b = \sum_{i=1}^4 a_i b_i$; and $Q_\mu$ spacelike $\Rightarrow$ $Q^2>0$.}

In connection with the elastic form factors:
\begin{equation}
P\cdot Q = 0\;\mbox{and}\; P^2 = -m_H^2-\frac{1}{4}Q^2 ,
\end{equation}
with $H=K,\pi$ as appropriate.  For the $K_{\ell 3}$ transitions, on the other hand:
\begin{subequations}
\label{Kl3kinematics}
\begin{eqnarray}
\label{Kl3kinematicsA}
2 P\cdot Q &=& m_K^2-m_\pi^2 =: \Delta_{K\pi}\,,\\
\label{Kl3kinematicsB}
2 P^2 &=& -(m_K^2+m_\pi^2)-\frac{1}{2}Q^2 =: -\Sigma_{K\pi} -\frac{1}{2}Q^2, \quad
\end{eqnarray}
\end{subequations}
and $t_m=(m_K-m_\pi)^2<\Delta_{K\pi}$ is the largest value of the squared-momentum-transfer in the physical decay process.  The kinematic constraints for the $\pi_{e 3}$ transition are obtained from Eqs.\,\eqref{Kl3kinematics} by the replacements $K\to \pi^+$, $\pi\to \pi^0$.

We judge it worth recalling that in the isospin symmetric limit, $f^\pi_+ \equiv - F_\pi$ and $f^\pi_- \equiv 0$.  Moreover, in the limit of exact $SU(3)$-flavour symmetry, $f^{\pi,K}_+ \equiv - F_\pi$ and $f^{\pi,K}_- \equiv 0$.  Consequently, $f_-^K$ should be a sensitive gauge of $SU(3)$-breaking. 
It is notable, too, that away from the $SU(3)$-flavour symmetry limit the Ademollo-Gatto theorem ensures $[f^{K}_+(0)]^2\approx 1$ \cite{Ademollo:1964sr}.  A consideration of contemporary theoretical estimates \cite{Sciascia:2008fr} indicates
\begin{equation}
\label{fp0emp}
|f^K_+(0)| =0.97\pm0.01\,.
\end{equation}
An experimental estimate can be inferred using values of the CKM matrix element $V_{us}$ in Ref.\,\cite{Beringer:1900zz}:
\begin{equation}
|f^K_+(0)| =0.9605 \pm 0.0061\,.
\end{equation}

As a measure of the divergence $Q_\mu M_\mu^{K_{\ell 3}}(P,Q)$, the function
\begin{equation}
f_0^K(Q^2) = f_+^K(Q^2) - \frac{Q^2}{m_K^2-m_\pi^2}\, f_-^K(Q^2)
\label{eq:f0definition}
\end{equation}
is also useful in characterising the transitions.  According to Ref.\,\cite{Callan:1966hu}, current algebra predicts
\begin{equation}
f_0^K(Q^2=-\Delta_{K\pi})= \frac{f_K}{f_\pi} +\Delta_{\rm CT}= 1.20 +\Delta_{\rm CT},
\label{f0Delta}
\end{equation}
where $\Delta_{K\pi}$, defined in Eq.\,\eqref{Kl3kinematicsA}, is an albeit unphysical momentum transfer, $f_{K}$ and $f_\pi$ are the mesons' leptonic decay constants \cite{Beringer:1900zz},
and $\Delta_{\rm CT} = {\rm O}(m_u,m_d)$.  The correction $\Delta_{\rm CT}$ is generally found to be small \cite{Cirigliano:2011ny} (namely, of a magnitude similar to the error in $f_K/f_\pi$) and is therefore neglected hereafter.

\subsection{Semileptonic decays}
In any study of hadron physics observables it is critical to preserve symmetries.   For example, if one does not ensure satisfaction of the vector Ward-Green-Takahashi identity \cite{Ward:1950xp,Green:1953te,Takahashi:1957xn} throughout a computation of the pion's elastic form factor; i.e., in the gap and Bethe-Salpeter equations, and in the expression for the matrix element in Eq.\,\eqref{pionFF}, one cannot even guarantee the pion will have unit charge \cite{Roberts:1994hh}.  The DSE framework is distinguished by the existence of at least two nonperturbative, symmetry-preserving truncation schemes \cite{Munczek:1994zz,Bender:1996bb,Chang:2009zb}.  Herein we use a truncation that may be described as leading-order in the scheme of Refs.\,\cite{Munczek:1994zz,Bender:1996bb}; namely, the rainbow-ladder approximation, because it is a quantitatively reliable tool for computation of the properties of pions and kaons, for reasons that are well understood \cite{Qin:2011xq,Chang:2012cc}.

In the rainbow-ladder treatment of a vector$\times$vector contact interaction, the matrix element in Eq.\,\eqref{Kl3FF} is
\begin{eqnarray}
\nonumber
M_\mu^{K_{\ell 3}}(P,Q) &=&
\surd 2\,  N_c\, {\rm tr}_{\rm D}\int \frac{d^4t}{(2\pi)^4} \Gamma_\pi(-p) S_u(t) \Gamma_K(k) \\ %
&& \times S_s(t+k)  i {\cal V}_{\mu}^{\,su}(Q) S_u(t+p)\,.
\label{ExplicitMKl3}
\end{eqnarray}
In Eq.\,\eqref{ExplicitMKl3}, $S_{u,s}$ are dressed-quark propagators and $\Gamma_{\pi,K}$ are meson Bethe-Salpeter amplitudes.  In the context of the rainbow-ladder truncation of the contact interaction, their forms are described in App.\,\ref{sec:contact}.

\subsubsection{Weak vector vertex}
\label{sec:WVV}
The remaining element in Eq.\,\eqref{ExplicitMKl3} is the vector piece of the dressed-quark--$W$-boson vertex, ${\cal V}_{\mu}^{su}(Q)$, which satisfies a Ward-Green-Takahashi identity:
\begin{equation}
\label{WGTI}
Q_{\mu} i {\cal V}^{su}_{\mu}(Q) = S_s^{-1}(q+Q) - S_u^{-1}(q)- (m_s-m_u)\Gamma^{su}_I(Q)\,,
%
\end{equation}
where $\Gamma_I^{su}$ is an analogous Dirac-scalar vertex.  (The axial-vector piece of the quark--$W$-boson coupling cannot contribute to a $0^- \to 0^-$ transition in the Standard Model.)  In order to highlight the symmetry-preserving nature of our treatment of the contact interaction, it is worth detailing the computation of ${\cal V}^{su}_{\mu}$.

Using the interaction kernels in App.\,\ref{sec:contact}, the inhomogeneous Bethe-Salpeter equation for the dressed-quark--W-boson vertex is
\begin{eqnarray}
\nonumber
\lefteqn{{\cal V}^{su}_\mu(Q) = \gamma_\mu} \\
&& - \frac{16\pi\alpha_{\rm IR}}{3 m_G^2}\int\! \frac{d^4t}{(2\pi)^4}
\gamma_\alpha S_s(t+Q) \Gamma^{su}_\mu(Q) S_u(t) \gamma_\alpha \label{vectorIBSE}
\end{eqnarray}
and that for its scalar counterpart is
\begin{eqnarray}
\nonumber
\lefteqn{\Gamma^{su}_I(Q) = I_{\rm D}} \\
&& - \frac{16 \pi\alpha_{\rm IR}}{3m_G^2}\int\! \frac{d^4t}{(2\pi)^4}
\gamma_\alpha S_s(t+Q) \Gamma^{su}_I(Q) S_u(t) \gamma_\alpha\,. \label{scalarIBSE}
\end{eqnarray}

With a symmetry-preserving treatment of the contact interaction, the vector vertex has the general form
\begin{equation}
\label{genvector}
{\cal V}^{su}_\mu(Q) = \gamma_\mu^T P_T^{su}(Q^2) + \gamma_\mu^L P_{1L}^{su}(Q^2) - i Q_\mu I_{\rm D} P_{2L}^{su}(Q^2)\,,
\end{equation}
where $Q_\mu \gamma_\mu^T = 0$, $\gamma_\mu^T+\gamma_\mu^L = \gamma_\mu$; and the scalar vertex may be written
\begin{equation}
\label{genscalar}
\Gamma^{su}_I(Q) = I_{\rm D} \, {\cal E}_I^{su}(Q^2) \,.
\end{equation}
These expressions are simple, in part because a momentum-independent interaction cannot support a dependence on relative momentum.

With these things in mind, consider Eq.\,\eqref{scalarIBSE}, which may be written
\begin{eqnarray}
\nonumber
\lefteqn{I_{\rm D} \, {\cal E}_I^{su}(Q^2) = I_{\rm D}}\\
&&
 - {\cal E}_I^{su}(Q^2) \frac{16\pi\alpha_{\rm IR}}{3m_G^2}\int\! \frac{d^4t}{(2\pi)^4}
\gamma_\alpha S_s(t+Q)  S_u(t) \gamma_\alpha\,.
\end{eqnarray}
It is straightforward to evaluate the expression on the second line: compute the Dirac contraction; introduce a Feynman parametrisation, characterised by the parameter $\alpha$; shift the integration variable $t\to t- \alpha Q$; and thereby arrive at
\begin{eqnarray}
\label{KEdefine}
\lefteqn{K_{\cal E}^{su}(Q^2)= \frac{16\pi\alpha_{\rm IR}}{3 m_G^2}\int\! \frac{d^4t}{(2\pi)^4}
\gamma_\alpha S_s(t+Q)  S_u(t) \gamma_\alpha}\quad \\
&=&  -\frac{4\alpha_{\rm IR}}{3\pi m_G^2} \int_0^1 d\alpha \int_0^\infty \! dy\, y
\frac{y - \alpha \hat\alpha Q^2-M_u M_{s}}{[y + w_{u\bar s}]^2}\,,\quad
\end{eqnarray}
where $w_{u\bar s}=\omega_{u\bar s}(\alpha,Q^2)$, with the latter defined in Eq.\,\eqref{eq:omega}.  At this point, Eqs.\,\eqref{eq:C0}, \eqref{eq:C1} may be used to obtain
\begin{eqnarray}
\nonumber
K_{\cal E}^{su}(Q^2) &= & -\frac{4\alpha_{\rm IR}}{3\pi m_G^2} \int_0^1 d\alpha \bigg[
{\cal C}^{\rm iu}(w_{u\bar s})
- {\cal C}^{\rm iu}_1(w_{u\bar s})\\
&& - (M_u M_s + \alpha \hat \alpha Q^2)\overline {\cal C}^{\rm iu}_1(w_{u\bar s})\bigg];
\label{KEsu}
\end{eqnarray}
and hence the scalar vertex is Eq.\,\eqref{genscalar} with
\begin{equation}
\label{answerScalar}
{\cal E}_I^{su}(Q^2) = \frac{1}{1+K_{\cal E}^{su}(Q^2)}\,.
\end{equation}
Comparison with Eq.\,(B15) in Ref.\,\cite{Chen:2012qr} shows that, as it should, the rainbow-ladder scalar vertex exhibits a pole at $Q^2=-m_\kappa^2$; i.e., at the mass of the lightest $u\bar s$-scalar excitation.

We return now to the vector vertex.  Substituting Eq.\,\eqref{genvector} into Eq.\,\eqref{vectorIBSE} and drawing upon Eq.\,\eqref{genscalar}, one finds
\begin{equation}
P_{1L}^{su}(Q^2) \equiv 1\,.
\end{equation}

One may determine $P_T^{su}(Q^2)$ by first contracting Eq.\,\eqref{vectorIBSE} with the transverse projection operator $T_{\mu\nu}(Q)=\delta_{\mu\nu}-Q_\mu Q_\nu/Q^2$, then proceeding as with the derivation of Eq.\,\eqref{answerScalar} and finally using Eq.\,\eqref{avwtiP}, to find
\begin{eqnarray}
\label{PTsu}
P_T^{su}(Q^2) &=& \frac{1}{1+K_V^{su}(Q^2)}\,,\\
\nonumber
K_V^{su}(Q^2)&=&-\frac{2\alpha_{\rm IR}}{3\pi m_G^2} \int_0^1 d\alpha \big[
M_u M_s - M_u^2 \hat\alpha \\
&& \quad - M_s^2\alpha-2 \alpha\hat\alpha Q^2\big]
\overline {\cal C}^{\rm iu}_1(w_{u\bar s})\,.
\end{eqnarray}
Comparison with Eq.\,(19) in Ref.\,\cite{Chen:2012qr} shows that, naturally, the transverse part of the rainbow-ladder vector vertex exhibits a pole at $Q^2=-m_{K^\ast}^2$; i.e., at the mass of the lightest $u\bar s$-vector excitation.

Only the computation of $P^{su}_{2L}(Q^2)$ remains.  This may be accomplished by first contracting Eq.\,\eqref{vectorIBSE} with $i Q_\mu$, then using the identity
\begin{equation}
i \gamma\cdot Q = S_s^{-1}(t+Q) - S_u^{-1}(t) -M_s + M_u
\end{equation}
and subsequently the gap equations for the $s,u$-quarks, Eq.\,\eqref{gap-1}, and finally Eq.\,\eqref{KEdefine}, so that one arrives at
\begin{equation}
Q^2 P^{su}_{2L}(Q^2) = (m_u-m_s) {\cal E}_I^{su}(Q^2) - M_u+M_s\,.
\end{equation}

It is straightforward now to verify the Ward-Green-Takahashi identity, Eq.\,\eqref{WGTI}, by direct substitution.

It is worth noting that ${\cal V}_\mu^{su}$ does not exhibit a pole at $Q^2=0$.  That is so because
\begin{equation}
(m_s-m_u) {\cal E}_I^{su}(Q^2=0) =  M_s-M_u\,,
\label{scalarWIdifference}
\end{equation}
as may readily be verified using the $u,s$-quark gap equations, Eq.\,\eqref{gap-1}, and Eqs.\,\eqref{genscalar}, \eqref{KEdefine}.

Equation\,\eqref{scalarWIdifference} is a particular example of a general identity that is true irrespective of the interaction.  In a renormalisable relativistic quantum gauge field theory the scalar vertex at zero total momentum, $Q=0$, takes the form
\begin{equation}
\Gamma_I^{su}(k;Q=0;\zeta) = I_{\rm D}{\cal E}_I^{su}(k^2;\zeta) + i \gamma\cdot k {\cal G}_I(k^2;\zeta)\,,
\end{equation}
where $k$ is the relative momentum and $\zeta$ is the renormalisation point.  The scalar functions in this expression satisfy
\begin{subequations}
\label{WGTIscalardifference}
\begin{eqnarray}
(m_s^\zeta-m_u^\zeta){\cal E}_I^{su}(k^2;\zeta) &=& B_s(k^2;\zeta) - B_u(k^2;\zeta)\,,\quad \\
(m_s^\zeta-m_u^\zeta){\cal G}_I^{su}(k^2;\zeta) &=& A_s(k^2;\zeta) - A_u(k^2;\zeta)\,,
\end{eqnarray}
\end{subequations}
where the general form of the dressed propagator for a fermion with flavour $f$ is given by
\begin{equation}
S_f^{-1}(k;\zeta) = i \gamma\cdot k A_f(k^2;\zeta) + I_{\rm D}\, B_f(k^2;\zeta)\,.
\end{equation}
Equations\,\eqref{WGTIscalardifference} will be recognised as finite-difference generalisations of better known scalar Ward identities \cite{Chang:2008ec}:
\begin{subequations}
\begin{eqnarray}
{\cal E}_I^{ff}(k^2;\zeta) &=& \frac{\partial}{\partial m^\zeta_f}B_f(k^2;\zeta)\,,\\
{\cal G}_I^{ff}(k^2;\zeta) &=& \frac{\partial}{\partial m^\zeta_f}A_f(k^2;\zeta)\,.
\end{eqnarray}
\end{subequations}

\subsubsection{Vertex nonanalyticities}
\label{sec:WVVnonA}
We have explained that, when computed in rainbow-ladder truncation, the vector vertex exhibits poles at the location of various bound-states.  For the $K_{\ell 3}$ transition form factors, this translates into the appearance of a pole in $f_+^K(Q^2)$ at $Q^2=-m_{K^\ast}^2$, where $m_{K^\ast}$ is the mass of the $K^\ast$-meson, and a pole in $f_0^K(Q^2)$ at $Q^2=-m_\kappa^2$, where $m_\kappa$ is the rainbow-ladder mass of the $u\bar s$-scalar-meson.

When one proceeds beyond rainbow-ladder truncation, these poles in the form factors are smeared by widths.  The same class of corrections to the rainbow-ladder truncation contains those diagrams whose point-meson analogue is $K \pi$ rescattering for the $K_{\ell 3}$ transition, and $KK$- and $\pi\pi$-rescattering for the respective elastic form factors.
Such effects are modest in the neighbourhood of $Q^2=0$.  For example, $\pi\pi$-rescattering increases the pion's charge radius by $\lesssim 10$\%, beyond that obtained in a complete rainbow-ladder treatment (which necessarily includes a simple-pole associated with $\rho$-meson) \cite{Alkofer:1993gu,Maris:1999bh,Roberts:2000aa,Roberts:2010rn}.  Furthermore, the impact of such corrections diminishes rapidly with increasing spacelike momenta because hard probes expose the structure of a hadron's dressed-quark core.

On the other hand, rescattering effects on form factors increase as the squared-momentum transfer moves toward a timelike value associated with a nonanalytic feature in the $S$-matrix, such as a production threshold.  This might be particularly important for the $K_{\ell 3}$ transitions because $t_m \approx (1/3) t_p$, $t_p=(m_K+m_\pi)^2$.  In Ref.\,\cite{Kalinovsky:1996ii} such considerations led to exploration of the impact of a beyond-rainbow-ladder correction to the weak vector vertex, which in our context corresponds to the following \emph{Ansatz}:
\begin{equation}
\label{VAnsatz}
{\cal V}_\mu^{su}(Q) \to {\cal V}_\mu^{su}(Q) + c_l \gamma_\mu {\rm e}^{t/t_p}{\cal L}^{su}(Q^2)\,,
\end{equation}
with $c_l$ a parameter, see Table\,\ref{Table:TransitionFF}, and
\begin{eqnarray}
\nonumber
{\cal L}^{su}(t=-Q^2) & = & 2+
\left[\frac{\Delta_{K\pi}}{t}-\frac{\Sigma_{K\pi}}{\Delta_{K\pi}}\right] \ln\frac{m_\pi^2}{m_K^2}\\
&&-\frac{\nu(t)}{t}\ln\frac{[t+\nu(t)]^2-\Delta_{K\pi}^2}{[t-\nu(t)]^2-\Delta_{K\pi}^2}\,,\quad
\label{clvertex}
\end{eqnarray}
where $\nu(t)^2=(t-t_p)(t-t_m)$.  It was shown in Ref.\,\cite{Kalinovsky:1996ii} that the precise form of ${\cal L}(t)$ is unimportant, only its analytic structure matters.  N.B.\,${\cal L}^{su}(t)$ is regular in the neighbourhood of $t=0$.

\subsubsection{Kaon leptonic transition form factors}
All elements necessary for the computation of $f_\pm^K(Q^2)$ are now in hand.  They may be obtained from $M(P,Q)$ in Eq.\,\eqref{ExplicitMKl3} as follows:
\begin{subequations}
\label{fpfmratios}
\begin{eqnarray}
f_+^K(Q^2) &= & \frac{1}{\surd 2}
\frac{Q^2 P\cdot M - P\cdot Q \,Q\cdot M}{P^2 Q^2 - (P\cdot Q)^2}\,,\\
f_-^K(Q^2)&=& \surd 2
\frac{P\cdot Q P\cdot M - P^2 \,Q\cdot M}{P^2 Q^2 - (P\cdot Q)^2}\,.
\end{eqnarray}
\end{subequations}
Formulae for the quantities $P\cdot M$ and $Q\cdot M$ are straightforward to compute but the expressions are lengthy and we choose not to reproduce them herein.

\subsubsection{Pion leptonic transition form factors}
The form factors associated with the $\pi_{e 3}$ decay may be obtained by following the pattern described above: one must merely change $\bar s \to \bar d$.

\subsection{Elastic}
\subsubsection{Kaon}
The matrix element in Eq.\,\eqref{Kelastic} can be decomposed thus:
\begin{equation}
M_\mu(P,Q) = e_u M_\mu^{uu\bar s}(P,Q) + e_{\bar s} M_\mu^{u\bar s\bar s}(P,Q)\,,
\end{equation}
where the expressions
\begin{subequations}
\begin{eqnarray}
M_\mu^{uu\bar s}(P,Q) &=& 2 P_\mu F_{K^+}^u(Q^2)\,, \label{FuKp}\\
M_\mu^{u\bar s\bar s}(P,Q) &=& 2 P_\mu F_{K^+}^{\bar s}(Q^2)
\end{eqnarray}
\end{subequations}
define the flavour-separated charged-kaon form factors such that
\begin{equation}
\label{eq:flavourseparated}
F_{K^+}(Q^2) = e_u \, F_{K^+}^u(Q^2) + e_{\bar s} \,F_{K^+}^{\bar s}(Q^2)\,.
\end{equation}
It is noteworthy that the canonical normalisation condition for the kaon Bethe-Salpeter amplitude ensures $F_{K^+}^u(Q^2=0)=1=F_{K^+}^{\bar s}(Q^2=0)$.  (See App.\,\ref{app:current} for additional comments.)

In rainbow-ladder truncation,
\begin{subequations}
\label{Mus}
\begin{eqnarray}
\nonumber M_\mu^{uu\bar s}(P,Q) &=& 2 N_c {\rm tr}_{\rm D}\! \int \frac{d^4 t}{(2\pi)^4}
i\Gamma_K(-p) S_u(t+p)\\
&& \times  i {\cal V}_\mu^{uu}(Q) S_u(t+k) i \Gamma_K(k) S_s(t) \,,
\label{MusA}\\
\nonumber M_\mu^{u\bar s\bar s}(P,Q) &=& 2 N_c {\rm tr}_{\rm D}\! \int \frac{d^4 t}{(2\pi)^4}
i\Gamma_K(-p) S_u(t) i\Gamma_K(k) \\
&& \times  S_s(t-k) i {\cal V}_\mu^{ss}(Q) S_s(t-p) \,. \label{MusB}
\end{eqnarray}
\end{subequations}
All elements in these expressions are already known.  In particular, the vertices ${\cal V}_\mu^{ff}(Q)$ are natural specialisations of Eq.\,\eqref{genvector}.  From this it is plain that, in rainbow-ladder truncation, the kaon electromagnetic form factor possesses poles at the masses of $\rho$- and $\phi$-like mesons.  The pole at the ground-state $\rho$-meson mass is naturally not kinematically accessible since $m_\rho < 2 m_K$.

The neutral kaon is not an eigenstate of the charge conjugation operation and hence this particle has a nonzero elastic form factor:
\begin{equation}
\label{eq:FK0Q2}
F_{K^0}(Q^2) = e_d \, F_{K^0}^d(Q^2) + e_{\bar s} \,F_{K^0}^{\bar s}(Q^2)\,,
\end{equation}
where
\begin{subequations}
\begin{eqnarray}
2 P_\mu F_{K^0}^d(Q^2) & = & M_\mu^{dd\bar s}(P,Q)  \,,\\
2 P_\mu F_{K^0}^{\bar s}(Q^2) &=& M_\mu^{d\bar s\bar s}(P,Q) \,,
\end{eqnarray}
\end{subequations}
with these last two expressions obtained by analogy with Eqs.\,\eqref{Mus}.
In the isospin symmetry limit, which we usually employ herein,
\begin{equation}
F_{K^0}^{d}(Q^2) = F_{K^+}^{u}(Q^2)\,,\;
F_{K^0}^{\bar s}(Q^2) = F_{K^+}^{\bar s}(Q^2)\,.
\end{equation}

Analysis of the expression in Eq.\,\eqref{FuKp} yields
\begin{eqnarray}
\nonumber \lefteqn{ F_{K^+}^u(Q^2) = P_T^{uu}(Q^2) \big[
T_{K,EE}^u(Q^2) E_{u\bar s}^2 \quad }\\
&& \quad  + T_{K,EF}^u(Q^2) E_{u\bar s} F_{u\bar s} + T_{K,FF}^u(Q^2) F_{u\bar s}^2 \big]\,,
\label{formulaFuKp}
\end{eqnarray}
where: the functions $T_{K}^u(Q^2)$ are given in App.\,\ref{app:fff}; $E_{u\bar s}$, $F_{u\bar s}$ are elements in the kaon's Bethe-Salpeter amplitude [Eq.\,\eqref{KaonBSA}]; and $P_T^{uu}(Q^2)$ is plain from Eq.\,\eqref{PTsu}.

It will be observed that Eq.\,\eqref{MusA} is mapped into Eq.\,\eqref{MusB} under the interchanges $\bar s\leftrightarrow u$, $k\leftrightarrow -p$.  The latter changes none of the kinematic relations.  Hence,
\begin{eqnarray}
\nonumber \lefteqn{ F_{K^+}^{\bar s}(Q^2) = P_T^{ss}(Q^2) \big[
T_{K,EE}^s(Q^2) E_{u\bar s}^2 \quad }\\
&& \quad  + T_{K,EF}^s(Q^2) E_{u\bar s} F_{u\bar s} + T_{K,FF}^s(Q^2) F_{u\bar s}^2 \big]\,,
\label{formulaFsKp}
\end{eqnarray}
with the functions $T_{K}^s(Q^2)$ obtained from the expressions in Eq.\,\eqref{eqs:TKEE} by the interchange $\bar s\leftrightarrow u$.

\subsubsection{Pion}
The expression for the charged-pion elastic form factor is obtained by setting $\bar s\to \bar d$ in the formulae described above.  Since we assume isospin symmetry, this process yields:
\begin{equation}
\label{eq:pionflavour}
F_{\pi^+}(Q^2) = (e_u+e_{\bar d}) \, F_{\pi^+}^u(Q^2) = F_{\pi^+}^u(Q^2)\,.
\end{equation}
It may also be read from Eq.\,(A1) in Ref.\,\cite{Roberts:2011wy}.

\section{Results - Elastic}
\label{sec:resultselastic}
All information necessary for computation of the form factors is now available.

\subsection{Charged kaon}
\label{subsec:chargedkaon}
\begin{figure}[t]
\centerline{\includegraphics[width=0.95\linewidth]{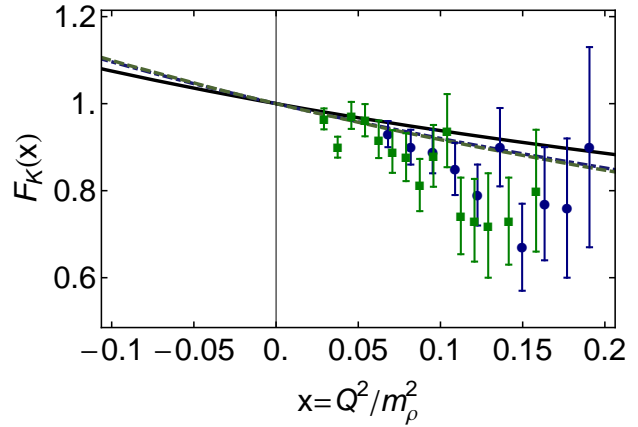}}
\centerline{\includegraphics[width=0.95\linewidth]{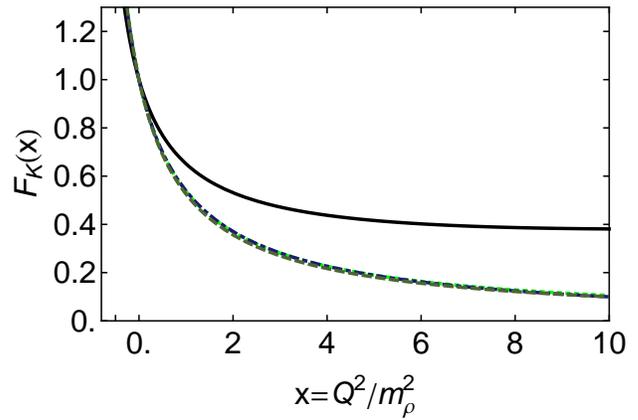}}
\caption{
\emph{Upper panel}.  Computed charged-kaon elastic form factor (solid curve) cf.\ the prediction in Ref.\,\protect\cite{Maris:2000sk} (dashed curve) and extant
data: circles \protect\cite{Dally:1980dj} and squares \protect\cite{Amendolia:1986ui}.
\emph{Lower panel}.  Comparison of the contact interaction result with the monopole fit  to the charged-kaon form factor in Ref.\,\protect\cite{Maris:2000sk}.
The dot-dashed curve in both panels is our computed result if one (erroneously) neglects the pseudovector component of the kaon; i.e., sets $F_{u\bar s}\equiv 0$ in Eq.\,\protect\eqref{KaonBSA}.
In all cases the results are rescaled with the appropriate value of $m_\rho$; namely, for the curves, that computed with the interaction, and for the data, the empirical value.
\label{fig:FKsmall}}
\end{figure}

We begin with the $K^+$ elastic form factor.  In the upper panel of Fig.\,\ref{fig:FKsmall} we depict the contact-interaction result for $F_K(Q^2)$, computed with the model parameters specified in Table~\ref{Tab:DressedQuarks} and Eq.\,\eqref{eq:theta}.  On the domain $x\in(-1,10]$, the function
\begin{equation}
F_K(x) \stackrel{\rm interpolation}{=}
\frac{1 +1.112 x + 0.228 x^2}{1+ 1.778 x + 0.780 x^2}\,,
\end{equation}
with $x=Q^2/m_\rho^2$, provides an accurate interpolation.  The figure also presents a comparison with both the DSE computation that successfully predicted the pion electromagnetic form factor \cite{Maris:2000sk} and extant data \cite{Dally:1980dj}.

Two things are immediately apparent.  First, as also observed elsewhere \cite{GutierrezGuerrero:2010md,Roberts:2010rn,Roberts:2011cf,Roberts:2011wy,Wilson:2011aa,Chen:2012qr}, in connection with observables determined by probes with $|Q^2|\lesssim M^2$, results obtained using a symmetry-preserving regularisation of the contact-interaction are not realistically distinguishable from those produced by the most sophisticated QCD renormalisation-group-improved kernels currently available.  In addition, available data on the charged-kaon form factor do not extend into the domain whereupon one could distinguish between contact-interaction results and those obtained with QCD-like kernels.

As the lower panel in Fig.\,\ref{fig:FKsmall} shows, the picture changes completely if one includes the domain $Q^2>M^2$.  It was demonstrated in Ref.\,\cite{Maris:1997hd} that pseudoscalar meson Bethe-Salpeter amplitudes necessarily possess components that may be described as pseudovector in character.  These components impact materially on a vast array of quantities involving pseudoscalar mesons and, of particular relevance herein, the large-$Q^2$ behaviour of their form factors \cite{GutierrezGuerrero:2010md,Maris:1998hc}.  Namely, if the meson is bound by an interaction whose behaviour at large relative momentum is $(1/k^2)^n$, then
\begin{equation}
\label{FMeson}
F_M(Q^2) \stackrel{Q^2 \gg M^2}{\simeq} \left[\frac{1}{Q^2}\right]^{n}\,,
\end{equation}
up to $\ln Q^2/M^2$ corrections, where $M$ is the infrared value of the dressed-quark mass. This explains the divergence between the solid and dashed curves in the lower panel of Fig.\,\ref{fig:FKsmall}: the former is obtained with our symmetry-preserving treatment of a contact interaction, $(1/k^2)^{n=0}$; and the latter with the like treatment of a QCD renormalisation-group-improved $(1/k^2)^{n=1}$ interaction.  The marked discrepancy highlights the potential for empirical data to chart the pointwise behaviour of the strong interaction between light-quarks.

In order to complete the illustration of these points, the dot-dashed curve in Fig.\,\ref{fig:FKsmall} depicts the result obtained with a contact interaction if the pseudovector component of the kaon's Bethe-Salpeter amplitude is erroneously omitted; i.e., one sets $F_{u\bar s}\equiv 0$ in Eq.\,\protect\eqref{KaonBSA}.  Plainly, if this error is made, as was formerly typical \cite{Blin:1987hw,Klevansky:1992qe}, then there is no realistic means by which to distinguish between $(1/k^2)^{n=0}$ and $(1/k^2)^{n=1}$ interactions, if the latter does not also make the mistake of ignoring the pseudoscalar meson's pseudovector components \cite{Roberts:1994hh}.  The dot-dashed curve in Fig.\,\ref{fig:FKsmall} may be interpolated using
\begin{equation}
F_{K}^{F\equiv 0}(x) \stackrel{\rm interp.}{=}
\frac{1 -0.237 x -0.749 x^2}{1 + 0.643 x -0.984 x^2 - 0.632 x^3 }.
\end{equation}

\begin{table}[t]
\caption{Row 1: Calculated radii, quoted in units of the computed value of $1/m_\rho$ (see Table.\,\ref{Tab:DressedQuarks}).  The ``$F\equiv 0$'' superscript indicates a result obtained by (erroneously) omitting the pseudovector components of the pseudoscalar mesons.  N.B.\,We quote the magnitude of the neutral kaon radius because $(r_K^0)^2 = -6 F^\prime_{K^0}(0)<0$.
Row 2: Results from Refs.\,\protect\cite{Maris:2000sk,Ji:2001pj}, expressed in units of that study's computed value of $1/m_\rho$.
Row 3: Experimental values \protect\cite{Beringer:1900zz}, quoted in terms of the empirical value of $1/m_\rho$.  These values span two columns since experiment cannot suppress a meson's pseudovector component.
N.B.\, $r_{K\pi}^2 = -6 f_+^{K\prime}(0)/f_+^K(0)$ and the empirical value is determined from the quadratic fit to the $K^\pm_{e3}$ form factor.
\label{Table:radii}
}
\begin{center}
\begin{tabular*}
{\hsize}
{
l|@{\extracolsep{0ptplus1fil}}
c@{\extracolsep{0ptplus1fil}}
c@{\extracolsep{0ptplus1fil}}
|c@{\extracolsep{0ptplus1fil}}
c@{\extracolsep{0ptplus1fil}}
|c@{\extracolsep{0ptplus1fil}}
c@{\extracolsep{0ptplus1fil}}
|c@{\extracolsep{0ptplus1fil}}
c@{\extracolsep{0ptplus1fil}}}\hline
       & $r_{K^+}$ & $r_{K^+}^{F\equiv 0}$
       & $|r_{K^0}|$ & $|r_{K^0}^{F\equiv 0}|$
       & $r_{\pi^+}$ & $r_{\pi^+}^{F\equiv 0}$
       & $r_{K\pi}$ & $r_{K\pi}^{F\equiv 0}$ \\\hline
computed  & $2.00$ & 2.30 & 0.78 & 0.85 & 2.12 & 2.40 & 1.79 & 2.10\\
\mbox{Ref.\,\protect\cite{Maris:2000sk,Ji:2001pj}}
          & $2.32$ & & $1.10$ & & 2.52 & & 2.44 & \\
empirical & \multicolumn{2}{c|}{$2.20 \pm 0.12$}
          & \multicolumn{2}{c|}{$1.09 \pm 0.07$}
          & \multicolumn{2}{c|}{$2.64 \pm 0.03$}
          & \multicolumn{2}{c}{$2.14 \pm 0.07$}\\  \hline
\end{tabular*}
\end{center}
\end{table}

Signficantly, the nearly identical behaviour of the dashed and dot-dashed curves was not achieved by fine tuning any parameters in our contact interaction: the same values were used for $F_{u\bar s}\neq 0$ as for $F_{u\bar s}\equiv 0$.  The curves are strikingly similar because the few parameters in both our interaction and that in Ref.\,\cite{Maris:2000sk} were fixed through fitting the same small set of static pion and kaon properties with equivalent accuracy.

In Table\,\ref{Table:radii} we list our calculated radii in comparison with both those computed in Ref.\,\cite{Maris:2000sk} and experiment \cite{Beringer:1900zz}.  The QCD renormalisation-group-improved interaction produces results in better agreement with empirical values.  However, analyses that (mistakenly) omit a pseudoscalar meson's pseudovector component are again seen to produce deceptively good results, something which explains, in part, the allure of such mistreatment of a contact interaction.

On the other hand, a more positive view might reasonably be advocated.  Bearing in mind that the veracious treatment of a contact-interaction is readily distinguishable from QCD, a $F_{f\bar g} \to 0$ \emph{Ansatz} may be used judiciously to produce a useful model of phenomena in hadron physics, so long as neither agreement nor disagreement with experiment is interpreted as a challenge to QCD.  Contemporary examples of this approach can be found in, e.g., Refs.\,\cite{Ito:2009zc,Cloet:2012td}; and we exploit it in Sec.\,\ref{subsec:pionplus}.

\begin{figure}[t]
\centerline{\includegraphics[width=0.95\linewidth]{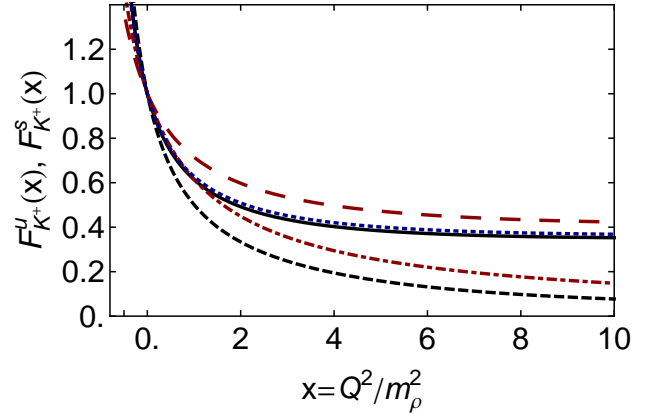}}
\caption{
Flavour-separated form factors for the $K^+$, defined in Eq.\,\eqref{eq:flavourseparated}.
Solid curve -- fully consistent contact interaction result for $F_{K^+}^u$;
short-dashed curve -- $F_{u\bar s}\equiv 0$ result for $F_{K^+}^u$;
long-dashed curve -- fully consistent contact interaction result for $F_{K^+}^{\bar s}$;
and dot-dashed curve -- $F_{u\bar s}\equiv 0$ result for $F_{K^+}^{\bar s}$.
For comparison, the dotted curve is $F_{\pi^+}^{u}=F_{\pi^+}^{\bar d}$ in Eq.\,\protect\eqref{eq:pionflavour}.
\label{fig:Fflavoursep}}
\end{figure}

\subsection{Flavour separated kaon elastic}
In Fig.\,\ref{fig:Fflavoursep} we depict the charged kaon's flavour separated form factors: $F_{K^+}^u$ and $F_{K^+}^{\bar s}$ in Eq.\,\eqref{eq:flavourseparated}.  The curves are accurately interpolated using
\begin{subequations}
{\allowdisplaybreaks
\begin{eqnarray}
F_{K^+}^u(x) &\stackrel{\rm interp.}{=}&
\frac{1 + 0.270 x + 0.0226 x^2}{1 + 1.050 x + 0.0541 x^2}\,,\\
F_{K^+}^{\bar s}(x) &\stackrel{\rm interp.}{=}&
\frac{1 + 0.221 x + 0.00893 x^2}{1 + 0.704 x + 0.0166 x^2}\,,\\
\nonumber
F_{K^+_{ F_{u\bar s}\equiv 0}}^{u}(x) &\stackrel{\rm interp.}{=}&
\frac{1 - 0.133 x + 0.0138 x^2}{1 + 0.859 x - 0.120 x^2+ 0.0160 x^3}\,,\\
&&\\
\nonumber
F_{K^+_{ F_{u\bar s}\equiv 0}}^{\bar s}(x) &\stackrel{\rm interp.}{=}&
\frac{1 + 0.227 x - 0.00538 x^2}{1 + 0.868 x + 0.121 x^2 - 0.00325 x^3}\,.\\
\end{eqnarray}}
\end{subequations}
One may readily compute radii from the interpolation formulae: in units of $1/m_\rho$, they are
\begin{equation}
\begin{array}{l|cc}
          & F_{u\bar s}\not\equiv 0 & F_{u\bar s}\equiv 0 \\\hline
r_{K^+}^u & 2.16 & 2.44 \\
r_{K^+}^{\bar s} & 1.70 & 1.96
\end{array}\,.
\end{equation}
For comparison, using Eq.\,\protect\eqref{eq:pionflavour}, one reads from Table\,\ref{Table:radii} that $r_{\pi^+}^u = 2.12$ or $2.40$.  This single measure highlights the picture painted in the figure; namely, identical to results obtained using a sophisticated interaction \cite{Maris:2000sk}, the dressed-$u$-quark charge distribution within the $K^+$ is almost indistinguishable from the dressed-$u$-quark charge distribution in the $\pi^+$.  In contrast, the result $r_{K^+}^{\bar s}/r_{K^+}^u \approx 0.8$ indicates, as one might have anticipated, that the heavier $\bar s$-quark is constrained to remain closer to the collective centre-of-mass within the charged kaon than the light $u$-quark.  It follows  that the $K^0$ elastic form factor, defined in Eq.\,\eqref{eq:FK0Q2}, should be positive on $Q^2>0$.

\begin{figure}[t]
\centerline{\includegraphics[width=0.95\linewidth]{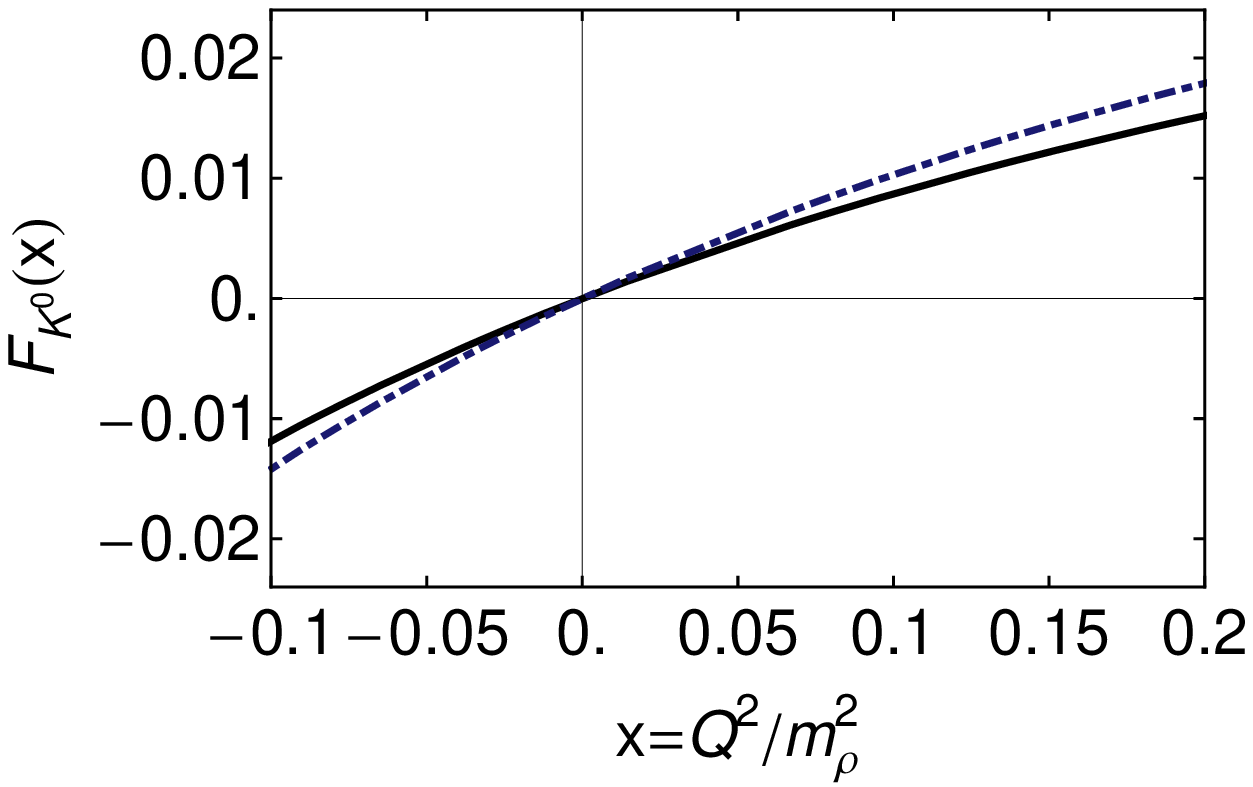}}
\centerline{\includegraphics[width=0.95\linewidth]{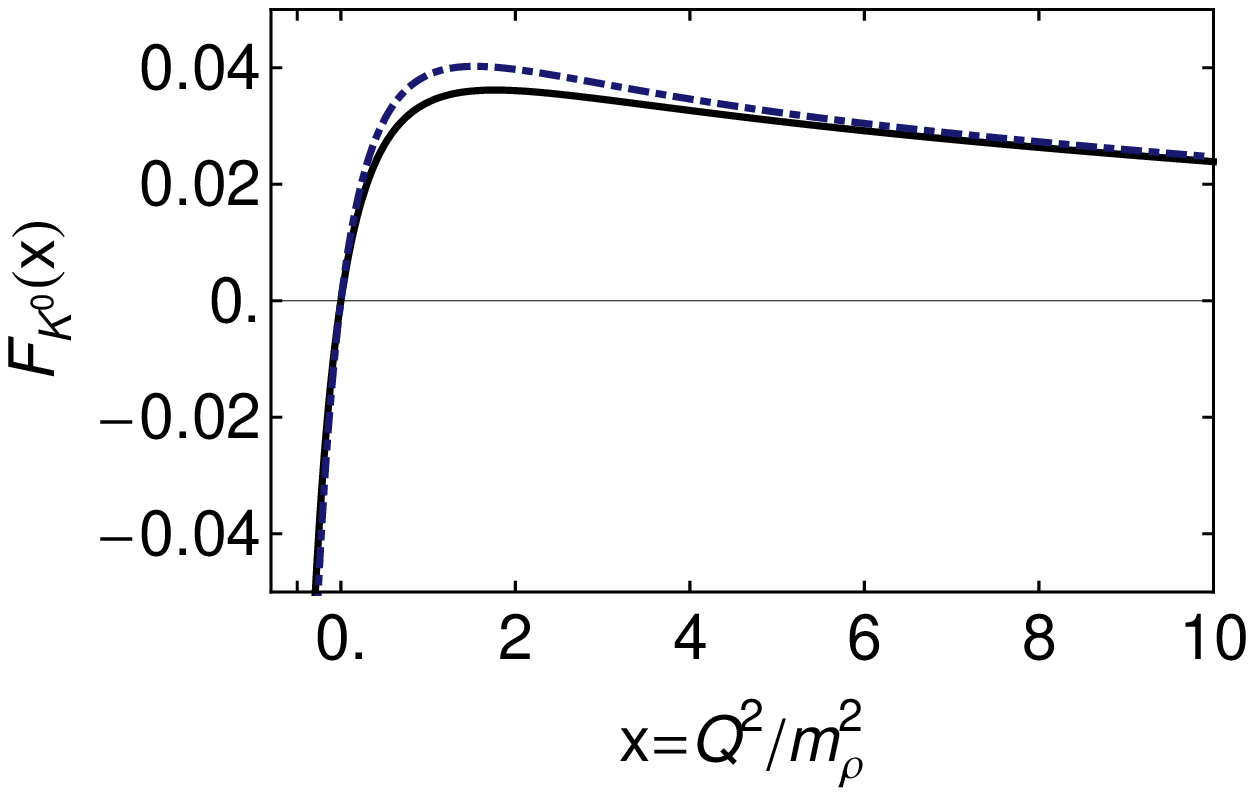}}
\caption{
Neutral kaon elastic form factor: solid curve -- complete Bethe-Salpeter amplitude; and dot-dashed curve -- result if one (erroneously) neglects the pseudovector component; i.e., sets $F_{u\bar s}\equiv 0$ in Eq.\,\protect\eqref{KaonBSA}.
\label{fig:FK0}}
\end{figure}

\subsection{Neutral kaon}
In Fig.\,\ref{fig:FK0} we depict the neutral kaon form factor.  In this case there is little difference on the displayed domain between the $F_{u\bar s}\neq 0$ and $F_{u\bar s}\equiv 0$ curves because that change has a nearly identical effect on both $F_{K^0}^d(Q^2)$ and $F_{K^0}^{\bar s}(Q^2)$, as apparent from Fig.\,\ref{fig:Fflavoursep}, and hence almost cancels in Eq.\,\eqref{eq:FK0Q2}.  The solid curve in these figures is accurately interpolated using
\begin{eqnarray}
\nonumber \lefteqn{F_{K^0}^{F\equiv 0}(x) \stackrel{\rm interp.}{=}}\\
&& \frac{x}{100} \frac{9.812 + 1.148 x + 0.0149 x^2 - 0.00346 x^3}
{1+1.667 x + 0.669 x^2}\,.
\end{eqnarray}

\subsection{Charged pion}
\label{subsec:pionplus}
The fully-consistent contact-interaction result for the charged pion's electromagnetic form factor is reported in Refs.\,\cite{GutierrezGuerrero:2010md,Roberts:2011wy}: naturally, one obtains [see Eq.\,\eqref{FMeson}]
\begin{equation}
F_{\pi^+}(Q^2) \stackrel{Q^2 \gg M^2}{\simeq} \mbox{constant}\,.
\end{equation}
We return to this topic here, however, in order to provide an illustration of the remarks at the end of Sec.\,\ref{subsec:chargedkaon}.

\begin{figure}[t]
\centerline{\includegraphics[width=0.95\linewidth]{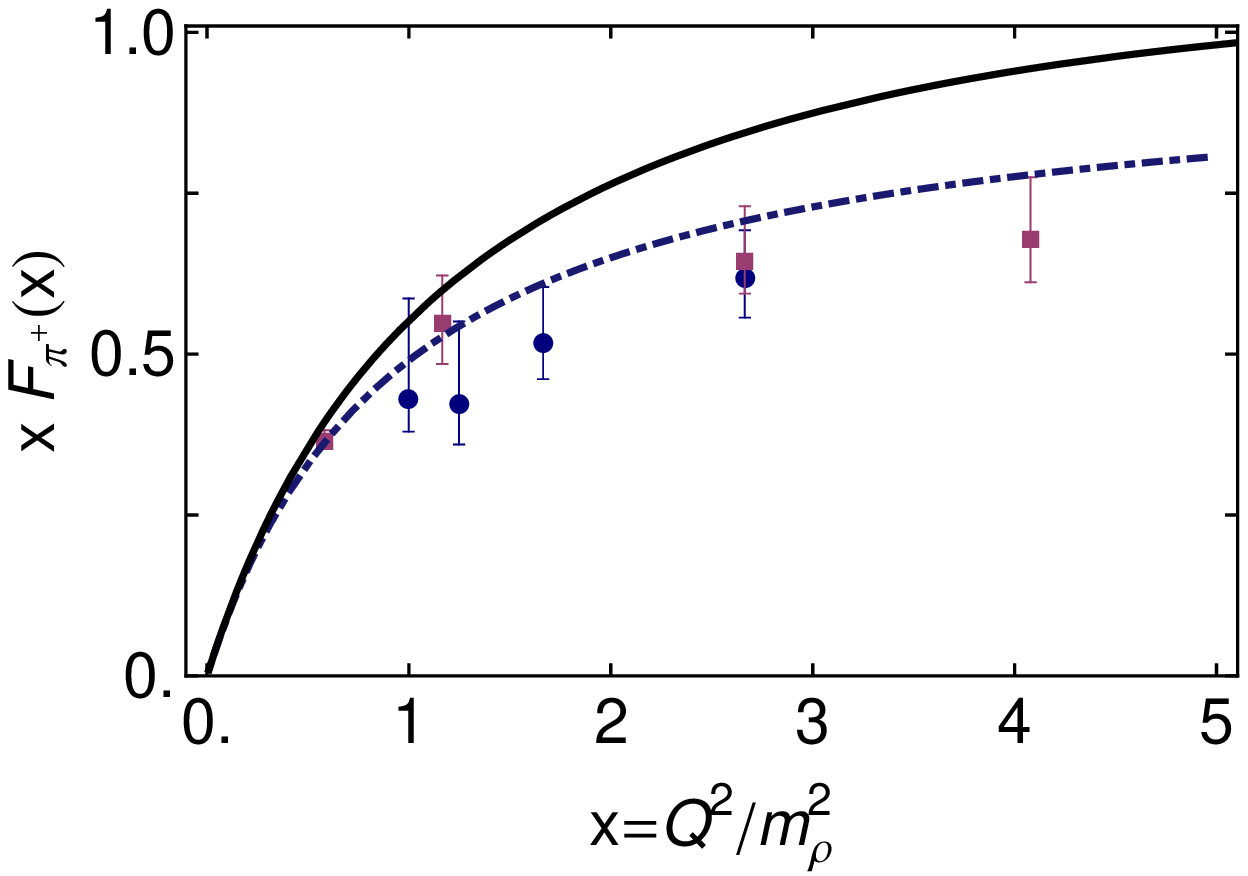}}
\centerline{\includegraphics[width=0.95\linewidth]{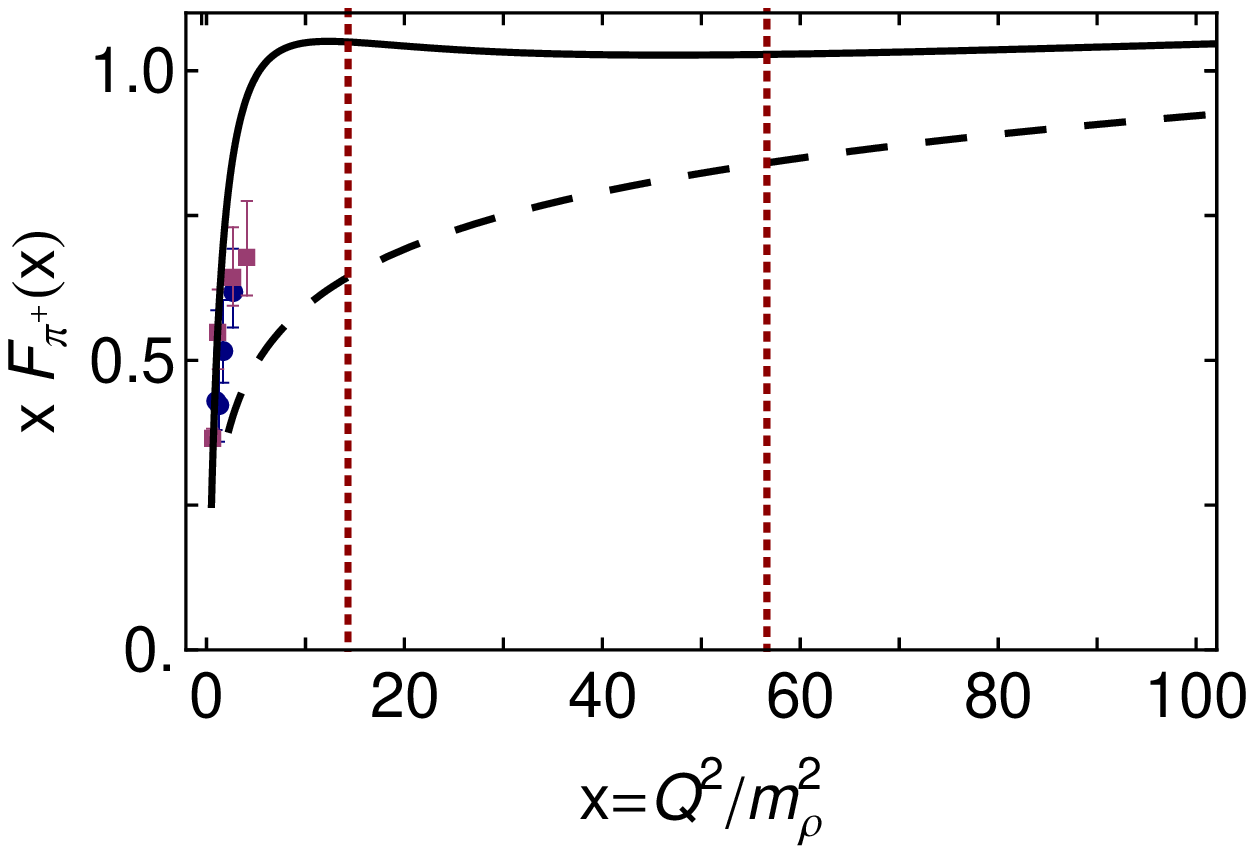}}
\caption{
Momentum-square weighted charged pion elastic form factor.
\emph{Upper panel}. Solid curve -- contact interaction result with $F_{u\bar d}\equiv 0$; and dot-dashed curve -- monopole interpolation of result in Ref.\,\protect\cite{Maris:2000sk}.  The data, included to illustrate the scale, are from Ref.\,\protect\cite{Huber:2008id}.
\emph{Lower panel}.  Solid curve -- contact interaction result with $F_{u\bar d}\equiv 0$; and long-dashed curve -- asymptotic form of the that result.  The vertical dotted lines at $x=14$ and $x=57$ mark, respectively, the points at which the asymptotic form is 60\% and 80\% of the full result.
\label{fig:Fpiplus}}
\end{figure}

In Fig.\,\ref{fig:Fpiplus} we present the result for $F_{\pi^+}(Q^2)$ obtained if the pseudovector component of the pion is deliberately suppressed.   In this case \cite{Blin:1987hw}:
\begin{equation}
F_{\pi^+}^{F_{u\bar d}\equiv 0}(Q^2) \stackrel{Q^2 \gg M^2}{\propto} \frac{\ln Q^2/M^2}{Q^2}\,.
\end{equation}
The power-law is the same as that predicted by QCD but the $\ln$-dependence is different \cite{Farrar:1979aw,Efremov:1979qk,Lepage:1980fj}.  Notwithstanding this, the mere presence of the $\ln$-term is useful for the illustration we wish to draw.

It is apparent from the upper panel of Fig.\,\ref{fig:Fpiplus} that the $F_{u\bar d}\equiv 0$ \emph{Ansatz} could readily be tuned to produce results in accord with contemporary data: there is little difference between this result and the calculation often cited in connection with the data \cite{Maris:2000sk}.  Moreover, some might argue that the data appear to be approaching a plateau but at a value above that one might simply infer from perturbative QCD.

The lower panel shows that the latter conclusion would probably be premature.  The solid curve is an interpolation of the $F_{u\bar d}\equiv 0$ result, accurate on $x\in (1,100]$; viz.,
\begin{equation}
 F_{\pi^+}^{F_{u\bar d}\equiv 0}(x) \stackrel{x > 1}{=}
\ln \left[x \frac{m_\rho^2}{M^2}\right]
\frac{1 + 0.0443 x}{1 + 2.506 x + 0.309 x^2}\,,
\end{equation}
and the dashed curve is the asymptotic limit of this formula:
\begin{equation}
x F_{\pi^+}^{F_{u\bar d}\equiv 0}(x) \stackrel{\mbox{asymp.}}{=}
0.143 \, \ln \left[x \frac{m_\rho^2}{M^2}\right] \,. \label{xFpiUV}
\end{equation}
The upper panel depicts the limit of available reliable data.  It covers a small domain, upon which calculations are evolving slowly.  It is evident in the lower panel that the $F_{u\bar d}\equiv 0$ \emph{Ansatz} result does not truly flatten until $x\gtrsim 10$, at which point the asymptotic limit of the formula is responsible for only half the magnitude.  The full curve is not even reasonably well approximated by Eq.\,\eqref{xFpiUV} until $x\gtrsim 50$.  (Empirically, this is $Q^2 \gtrsim 30\,$GeV$^2$.)  It is at approximately $x=50$ that the curve exhibits a local minimum, a feature which marks the beginning of the domain whereupon the $\ln x$ growth is actually visible.  (In QCD, that would be a $\ln x$ suppression, which could be even harder to distinguish.)

We do not pretend that these observations amount to a statement about the domain upon which one might finally expect to discover perturbative QCD behaviour of the charged pion form factor.  (Note, on the other hand, that they are consistent with the conclusions drawn using a momentum-dependent dressed-quark mass \cite{Maris:1997hd}.)  It does, however, demonstrate concretely that: the approach to an asymptotic limit which involves logarithmic evolution can be very slow; and a plateau that seems to appear on a small domain can easily be misleading.

\section{Results - Transition}
\label{sec:resultstransition}

We turn now to the $K_{\ell 3}$ transition form factors, the physical domain for which is $t\in [m_\ell^2,t_m]$.  

\begin{figure}[t]
\centerline{\includegraphics[width=0.95\linewidth]{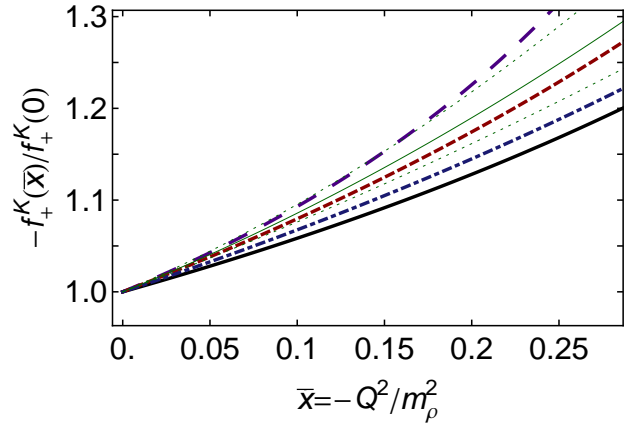}}
\centerline{\includegraphics[width=0.95\linewidth]{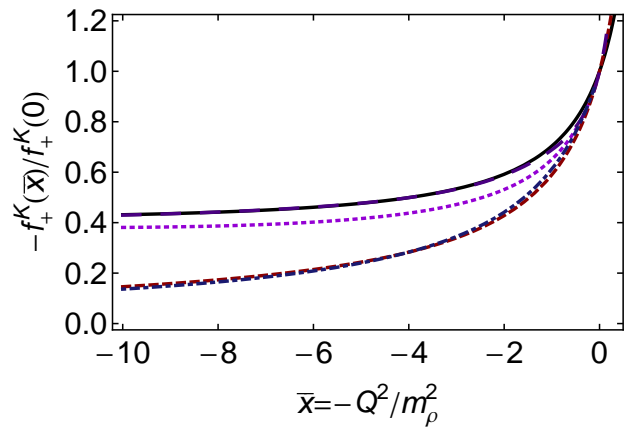}}
\caption{
$K_{\ell 3}$ transition form factor $f_+^K$.
\emph{Upper panel}.  Solid curve -- contact interaction; dashed curve -- contact interaction with elimination of the pseudovector component in both pseudoscalar mesons ($F_{0^-}\to 0$); long-dashed curve -- contact interaction, obtained with vertex in Eq.\,\protect\eqref{VAnsatz}; thin curve bracketed by dotted curves -- quadratic fit to empirical data, drawn from Ref.\,\protect\cite{Beringer:1900zz}; and dot-dashed curve -- result from Ref.\,\protect\cite{Ji:2001pj}.
%
%
\emph{Lower panel}.  Solid curve -- contact interaction; dashed curve -- contact interaction with elimination of the pseudovector component in both pseudoscalar mesons ($F_{0^-}\to 0$); long-dashed curve -- contact interaction, obtained with vertex in Eq.\,\protect\eqref{VAnsatz}; dot-dashed curve -- monopole interpolation of the result from Ref.\,\protect\cite{Ji:2001pj}; and dotted curve -- elastic kaon form factor, which is the solid curve in the lower panel of Fig.\,\protect\ref{fig:FKsmall}.
\label{fig:FKl3p}}
\end{figure}

\subsection{Primary kaon transition form factor}
Our result for $f_+^K$, depicted in Fig.\,\ref{fig:FKl3p}, is accurately interpolated by
%
\begin{equation}
f_+^{K}(x) \stackrel{\rm interp.}{=}
f_+^{K}(0) \frac{ 1 -4.492 x + 1.131 x^2 - 0.00966 x^3}{1 -5.029 x + 3.397 x^2}\,,
\end{equation}
with $f_+^{K}(0)$ listed in Table\;\ref{Table:TransitionFF}.
As seen with the elastic form factors and evident in the lower panel of Fig.\,\ref{fig:FKl3p}, the contact interaction result for $f_+^K$ (solid curve) is hard.

Omitting (erroneously) the pseudovector components in both pseudoscalar mesons, a much softer result is obtained.  This is the dashed curve in Fig.\,\ref{fig:FKl3p}, which is interpolated by
\begin{eqnarray}
\nonumber \lefteqn{f_+^{K_{F\equiv 0}}(x) \stackrel{\rm interp.}{=}}\\
&& f_+^{K_{F\equiv 0}}(0)
\frac{1+0.763 x - 0.203 x^2}{1+ 0.0299 x - 0.808 x^2 + 0.105 x^3}\,.
\end{eqnarray}

The long-dashed curve in Fig.\,\ref{fig:FKl3p}, which exhibits the steepest ascent, is produced by augmenting the contact-interaction vertex with the \emph{Ansatz} in Eq.\,\eqref{clvertex} that models a nonanalyticity associated with the $K\pi$ production threshold.  The parameter value $c_l=0.095$ was chosen in order to satisfy Eq.\,\protect\eqref{f0Delta}.  The lower panel shows that such nonanalyticities rapidly become immaterial when considering the evolution of form factors into the domain $Q^2>0$ \cite{Alkofer:1993gu}.  A valid interpolation of the long-dashed curve is provided by
\begin{eqnarray}
\nonumber \lefteqn{f_+^{K_{\mbox{\footnotesize Eq.\,\protect\eqref{VAnsatz}}}}(x)
\stackrel{\rm interp.}{=} }\\
&&
f_+^{K_{\mbox{\footnotesize Eq.\,\protect\eqref{VAnsatz}}}}(0) \frac{1-0.428 x + 0.450 x^2 - 0.264 x^3}{1-1.216 x+0.196 x^2-0.699x^3}\,.
\end{eqnarray}

The final calculation depicted in the upper panel of Fig.\,\ref{fig:FKl3p} is that from Ref.\,\cite{Ji:2001pj} (dot-dashed curve), which employed one-loop QCD renormalisation-group-improved kernels for the gap and Bethe-Salpeter equations.  The vertex computed therein possesses a pole at $t=(m_K^\ast)^2$ but no other nonanalyticities; and that is the origin of the difference between this result and the long-dashed curve.  The lower panel highlights features we have already seen with the elastic form factor.  The dot-dashed and dashed curves are nearly identical in the spacelike region; and hence we see again that a deliberate mistreatment of the contact interaction, by neglecting pseudovector components of pseudoscalar mesons, produces results that are not practically distinguishable from the those obtained with more sophisticated interactions.


\begin{figure}[t]
\centerline{\includegraphics[width=0.95\linewidth]{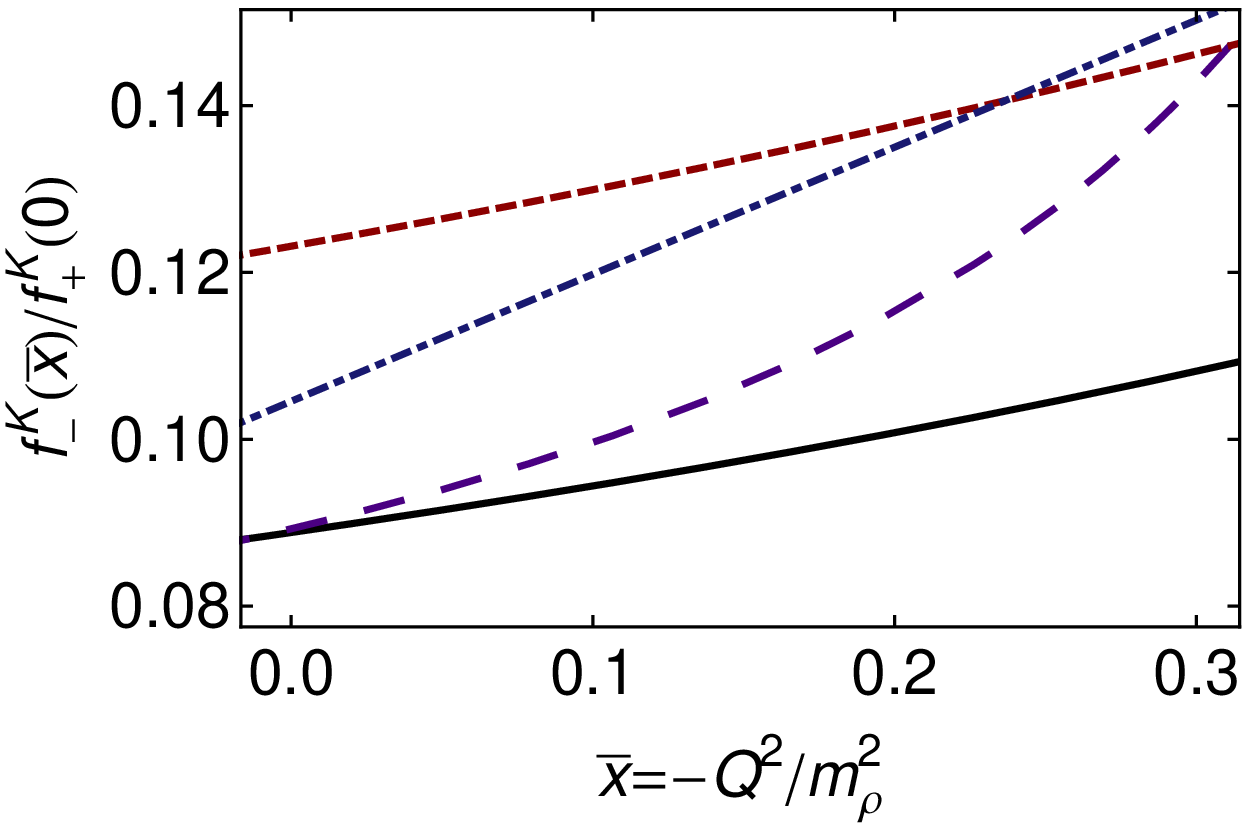}}
\centerline{\includegraphics[width=0.95\linewidth]{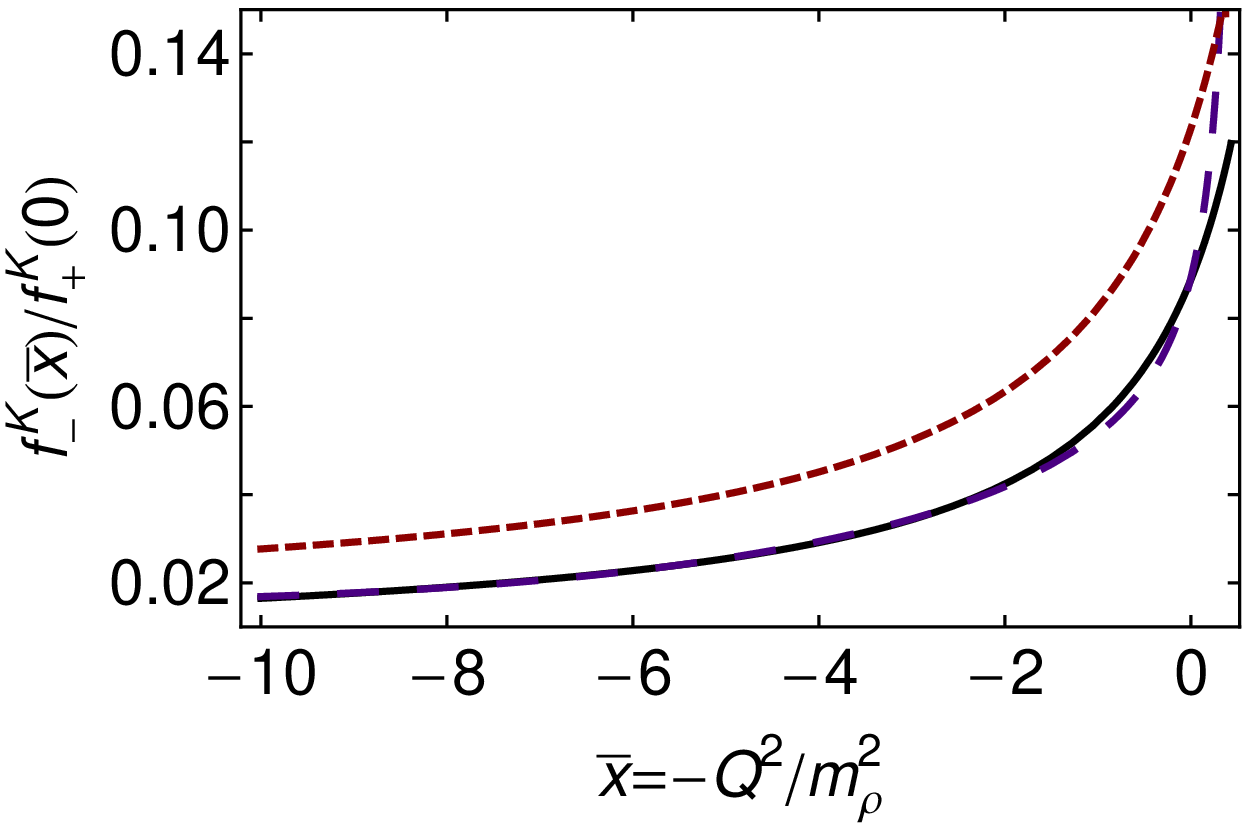}}
\caption{
$K_{\ell 3}$ transition form factor $f_-^K$.
\emph{Upper panel}.  Solid curve -- contact interaction; dashed curve -- contact interaction with elimination of the pseudovector component in both pseudoscalar mesons ($F_{0^-}\to 0$);  long-dashed curve -- contact interaction, obtained with vertex in Eq.\,\protect\eqref{VAnsatz}; and dot-dashed curve -- result extracted from Ref.\,\protect\cite{Ji:2001pj} on the domain within which it is valid.
\emph{Lower panel}.  Legend unchanged but the domain is extended.
\label{fig:FKl3m}}
\end{figure}

Curiously, the best agreement with the quadratic fit to empirical data (drawn from Ref.\,\protect\cite{Beringer:1900zz}) is obtained via mistreating the contact interaction by neglecting the pseudovector components of the pseudoscalar mesons.  Realistically, though, the data is incapable of distinguishing between the models.  On the other hand, much more care would be needed in formulating the model interactions before they could be useful in precision kaon physics.

In the lower panel of Fig.\,\ref{fig:FKl3p}, the charged-kaon elastic form factor is plotted as the dotted curve.  The difference between this and the solid curve is one measure of the magnitude of $SU(3)$-flavour symmetry breaking in our symmetry-preserving treatment of the contact interaction.  The breaking is observable but not dramatic.

\subsection{Secondary kaon transition form factor}
We plot $f_-^K$ in Fig.\,\ref{fig:FKl3m}.  This form factor may be interpolated using
\begin{equation}
f_-^{K}(x) \stackrel{\rm interp.}{=}
f_-^{K}(0)
\frac{1- 0.914 x + 0.0328 x^2}{1-10.879 x+6.403 x^2}\,.
\end{equation}
For comparison, we depict a result extracted from Ref.\,\cite{Ji:2001pj} (dot-dashed curve) on the domain within which it is valid: the scale and evolution rate are similar to those of our results obtained in the absence of the $K\pi$ threshold correction to the vertex, Eq.\,\eqref{VAnsatz}.

With respect to the other curves, there is one novelty in the pattern of comparison, evident in the lower panel: for this subdominant form factor the result obtained by neglecting the pseudovector components of the pseudoscalar mesons is not noticeably softer than that produced by a consistent treatment of the contact interaction.
This is readily understood upon careful consideration of Eqs.\,\eqref{Kl3kinematics}, \eqref{genvector}, \eqref{KEsu}, \eqref{answerScalar}, \eqref{fpfmratios}.
The $P\cdot M$ term in Eq.\,\eqref{fpfmratios} is sensitive to the presence or absence of $F_{u\bar s}$, whereas $Q\cdot M$ is not because it is dominated by the scalar vertex; and, owing to kinematics, Eqs.\,\eqref{Kl3kinematics}, $Q\cdot M$ dominates $f_-$ at large $Q^2$.

The dashed curve in Fig.\,\ref{fig:FKl3m} may be interpolated using
\begin{eqnarray}
\nonumber \lefteqn{f_-^{K_{F\equiv 0}}(x) \stackrel{\rm interp.}{=}}\\
&& f_-^{K_{F\equiv 0}}(0)
\frac{1+0.763 x - 0.203 x^2}{1+ 0.0299 x - 0.808 x^2 + 0.105 x^3}
\end{eqnarray}
and the long-dashed curve with
\begin{eqnarray}
\nonumber \lefteqn{f_-^{K_{\mbox{\footnotesize Eq.\,\protect\eqref{VAnsatz}}}}(x)
\stackrel{\rm interp.}{=} }\\
&&
f_-^{K_{\mbox{\footnotesize Eq.\,\protect\eqref{VAnsatz}}}}(0) \frac{1-0.428 x + 0.450 x^2 - 0.264 x^3}{1-1.216 x+0.196 x^2-0.699x^3}\,.
\end{eqnarray}

We recall now the observation made before Eq.\,\eqref{fp0emp}; namely, that $f_-^K$ should be a sensitive gauge of $SU(3)$-flavour symmetry breaking.  Observe, therefore, that
\begin{equation}
\label{fminusK0}
f_-^{K}(0) \approx 0.1
\approx 0.6 \frac{m_s-m_u}{\Lambda_{\rm uv}}
\approx 0.6 \frac{M_s-M_u}{\Lambda_{\rm uv}}\,;
\end{equation}
and so the $t=0$ value of this form factor is truly a direct measure of the symmetry breaking.  It is noteworthy that the magnitude found here is similar to the amount of $SU(3)$-flavour symmetry breaking observed in the difference between $\rho K K$ and $\rho\pi\pi$ couplings \cite{ElBennich:2011py}.

\begin{figure}[t]
\centerline{\includegraphics[width=0.95\linewidth]{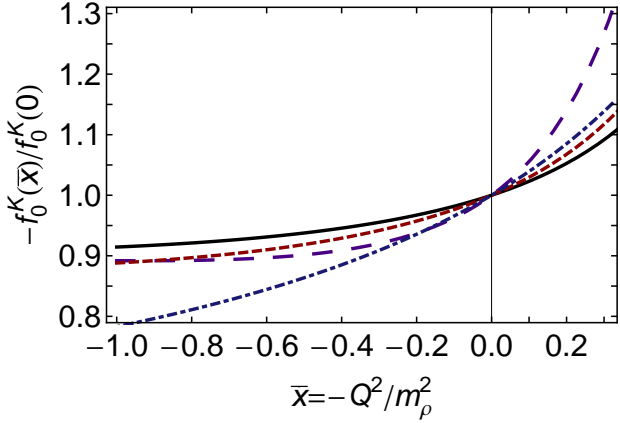}}
\caption{
$K_{\ell 3}$ transition form factor $f_0^K$.
Solid curve -- contact interaction; dashed curve -- contact interaction with elimination of the pseudovector component in both pseudoscalar mesons ($F_{0^-}\to 0$); long-dashed curve -- contact interaction, obtained with vertex in Eq.\,\protect\eqref{VAnsatz}; and dot-dashed curve -- result from Ref.\,\protect\cite{Ji:2001pj}.
\label{fig:FKl30}}
\end{figure}

\subsection{Kaon longitudinal transition form factor}
In Fig.\,\ref{fig:FKl30} we plot the $K_{\ell 3}$ scalar form factor, defined in Eq.\,\eqref{eq:f0definition}.  The solid curve may be interpolated using
\begin{eqnarray}
\nonumber\lefteqn{f_0^{K}(x) \stackrel{\rm interp.}{=}}\\
&& f_0^{K}(0) \frac{1-0.784 x + 0.0178 x^2+0.00113 x^3}{1-0.987 x - 0.0184 x^2}\,.
\end{eqnarray}
%
The pattern of comparison between the results should now be familiar.  An interpolation of the dashed curve is provided by
\begin{eqnarray}
\nonumber \lefteqn{f_0^{K_{F\equiv 0}}(x) \stackrel{\rm interp.}{=}}\\
&& f_0^{K_{F\equiv 0}}(0)
\frac{1-2.290  x + 1.181 x^2 - 0.0647 x^3}{1+ 2.553 x + 1.554 x^2}
\end{eqnarray}
%
%
and, of the long-dashed curve, by
\begin{eqnarray}
\nonumber \lefteqn{f_0^{K_{\mbox{\footnotesize Eq.\,\protect\eqref{VAnsatz}}}}(x)
\stackrel{\rm interp.}{=} }\\
&&
f_0^{K_{\mbox{\footnotesize Eq.\,\protect\eqref{VAnsatz}}}}(0)
\frac{1-1.372 x + 0.984 x^2+0.0261 x^3}{1-1.825 x - 0.909 x^2}\,.
\end{eqnarray}

\begin{table}[t]
\caption{A range of quantities that are typically used to characterise the semileptonic decays of pseudoscalar mesons.
Column 1: Results computed using vertex obtained as solution of Eq.\,\protect\eqref{vectorIBSE}.
Column 2: Results computed with solution of Eq.\,\protect\eqref{vectorIBSE} augmented by the \emph{Ansatz} in Eq.\,\eqref{VAnsatz}.  The parameter $c_l=0.095$ was chosen in order to produce $f_0^K(-\Delta)=1.20$ [see Eq.\,\protect\eqref{f0Delta}].
Column 3: Results computed using vertex obtained as solution of Eq.\,\protect\eqref{vectorIBSE} and replacement $F_{u\bar s}\to 0$.
Column 4: Results for comparable quantities reported in Ref.\,\protect\cite{Ji:2001pj}.
Column 5: Some empirical values inferred using Refs.\,\protect\cite{Sciascia:2008fr,Cirigliano:2011ny,Beringer:1900zz}: the widths correspond to $7.926 \pm 0.032 \times 10^6/s$ and $5.285 \pm 0.022 \times 10^6/s$.
The $\lambda$-parameters are defined in Eqs.\,\protect\eqref{lambda1}, \protect\eqref{lambda2}.
\label{Table:TransitionFF}
}
\begin{center}
\begin{tabular*}
{\hsize}
{
c|@{\extracolsep{0ptplus1fil}}
c@{\extracolsep{0ptplus1fil}}
c@{\extracolsep{0ptplus1fil}}
c@{\extracolsep{0ptplus1fil}}
|c@{\extracolsep{0ptplus1fil}}
|c@{\extracolsep{0ptplus1fil}}}\hline
       & \protect\eqref{vectorIBSE}
       & (\protect\ref{vectorIBSE},\protect\ref{VAnsatz})
       & $F_{f \bar g}\equiv 0\;$ & Ref.\,\protect\cite{Ji:2001pj}$\;$
       & Emp.\ \protect\cite{Sciascia:2008fr,Cirigliano:2011ny,Beringer:1900zz} \\\hline
$-f_+^K(-t_m)$ & 1.07\phantom{9} & 1.13\phantom{9} & 1.11\phantom{9} & 1.13 & $1.161 \pm 0.031$\\
$-f_+^K(0)$    & 0.98\phantom{9} & 0.98\phantom{9}  & 0.98\phantom{9} & 0.96 & $0.961 \pm 0.006$\\
$ f_-^K(-t_m)$ & 0.096 & 0.10\phantom{9} & 0.13\phantom{9}  & 0.11 & \\
$ f_-^K(0)$    & 0.087 & 0.088 & 0.12\phantom{9} & 0.10 & $0.120 \pm 0.023$ \\
$-f_0^K(-\Delta)$ & 1.06\phantom{9} & 1.20\phantom{9} & 1.08\phantom{9} & 1.18 & \\
$100\,\lambda^\prime_{K^\pm_{e3}}$
        & 1.21\phantom{5} & 1.78\phantom{9} & 1.66\phantom{5} & 2.23 & $2.485\pm 0.167$ \\
$100\,\lambda^{\prime\prime}_{K^\pm_{e3}}$
        & 0.044           & 0.12\phantom{9} & 0.060           & 0.10  & $0.192 \pm 0.094$ \\
$10\,\tilde \lambda^\prime_{K^\pm_{e3}}$
        & 5.37\phantom{5} & 7.88\phantom{9} & 7.33\phantom{5} & 6.33 & $7.667 \pm 0.513$ \\
$\phantom{10}\tilde \lambda^{\prime\prime}_{K^\pm_{e3}}$
        & 0.87\phantom{5}     & 2.43\phantom{9} & 1.17\phantom{5} & 0.80  & $1.825 \pm 0.896$ \\
$10^{18}\Gamma_{Ke3}/m_\rho$ & 5.53 & 5.66 & 5.62 & 6.54 & $6.721 \pm 0.027$ \\
$10^{18}\Gamma_{K\mu3}/m_\rho$ & 3.61 & 3.76 & 3.68 & 4.34 & $4.482 \pm 0.018$ \\\hline
\end{tabular*}
\end{center}
\end{table}

In Table\,\ref{Table:TransitionFF} we list a few quantities that are typically used to characterise the semileptonic decays of pseudoscalar mesons.  In this table, the usual slope parameters are defined via
\begin{subequations}
\label{lambda1}
\begin{eqnarray}
\lambda^\prime_{K^\pm_{e3}} &=& \left. m_{\pi^+}^2 \frac{d}{dt} f_+^{K_{e3}}(t)\right|_{t=0},\\
\lambda^{\prime\prime}_{K^\pm_{e3}} &= &
\left. m_{\pi^+}^4 \frac{d^2}{dt^2} f_+^{K_{e3}}(t)\right|_{t=0}.
\end{eqnarray}
\end{subequations}
In comparing them with the figures, it is important to recall that the $x$-axis for each curve is rescaled by the appropriate value of $m_\rho$.  The following slope parameters account for this:
\begin{equation}
\label{lambda2}
\tilde \lambda^\prime_{K^\pm_{e3}} = \frac{m_\rho^2}{m_\pi^2}\lambda^\prime_{K^\pm_{e3}}\,,\;
\tilde \lambda^{\prime\prime}_{K^\pm_{e3}} =\frac{m_\rho^4}{m_\pi^4}\lambda^{\prime\prime}_{K^\pm_{e3}} .
\end{equation}

The table includes widths for the neutral-kaon leptonic decays, computed using Eq.\,(11) in Ref.\,\cite{Nam:2007fx}, which corrects a typographical error in Eq.\,(14) of Ref.\,\cite{Ji:2001pj}.  As usual, the results are scaled by the appropriate value of the $\rho$-meson mass.

\begin{figure}[t]
\centerline{\includegraphics[width=0.95\linewidth]{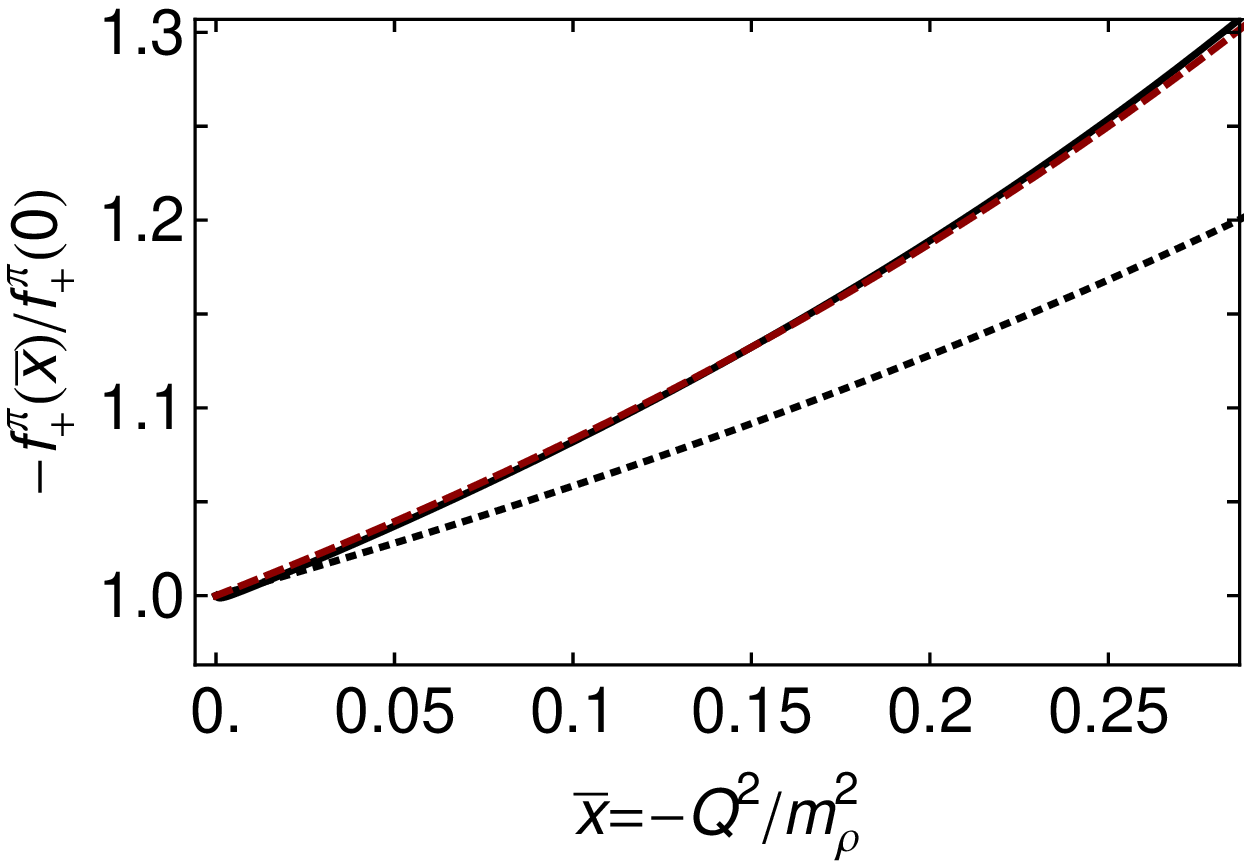}}
\centerline{\includegraphics[width=0.95\linewidth]{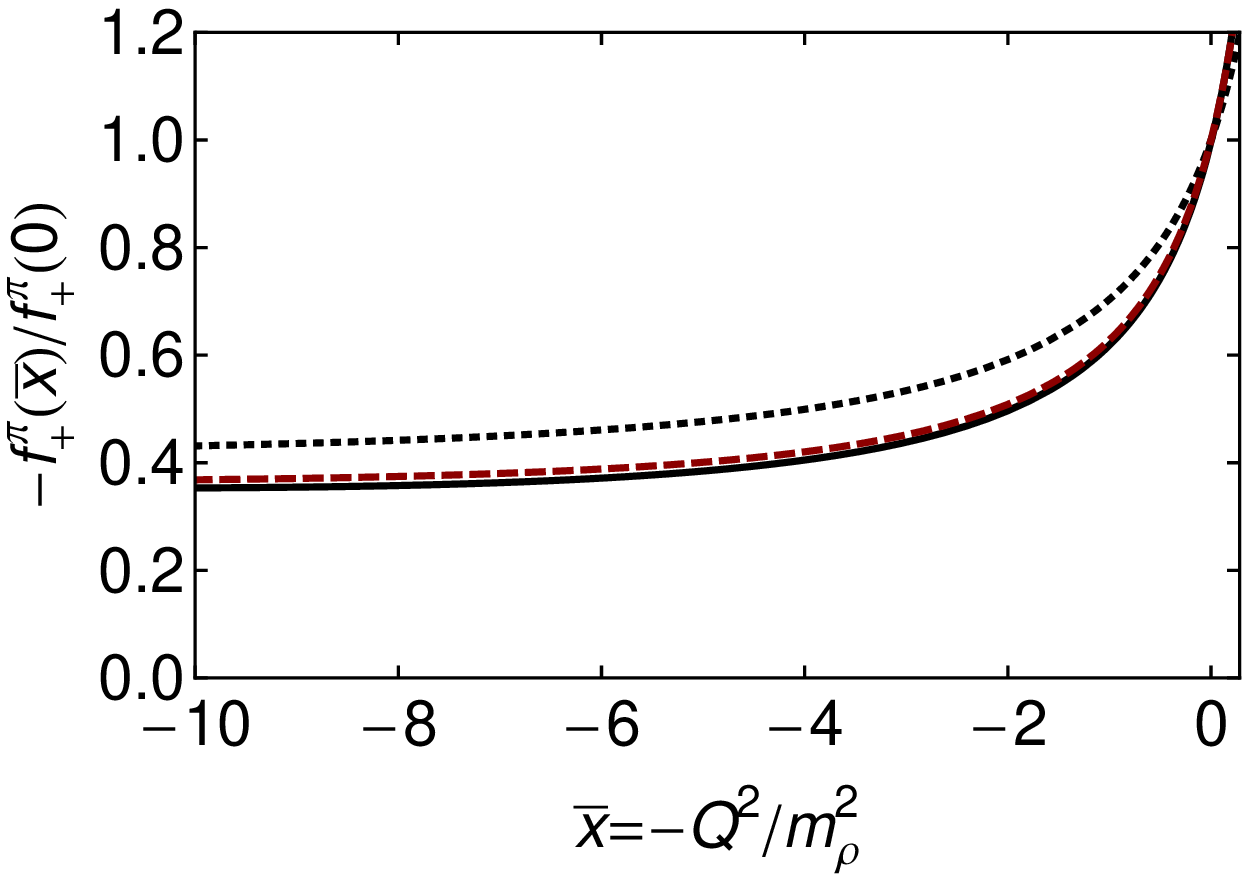}}
\caption{
$\pi_{e 3}$ transition form factor $f_+^\pi$.
\emph{Upper panel}.  Solid curve -- contact interaction; dashed curve -- interpolation of the pion elastic form factor from Ref.\,\protect\cite{Roberts:2011wy}; and dotted curve -- $f_+^K$ from Fig.\,\protect\ref{fig:FKl3p}.
\emph{Lower panel}.  Legend unchanged but the domain is extended.
\label{fig:Fpiep}}
\end{figure}

\subsection{Pion transition}
We depict $f_+^\pi$ in Fig.\,\ref{fig:Fpiep}: $f_+^\pi(0)=1.00$.  This form factor is almost identical to the pion elastic form factor, as ought to be the case; and deviates by a measurable but modest amount from $f_+^K$.

\begin{figure}[t]
\centerline{\includegraphics[width=0.95\linewidth]{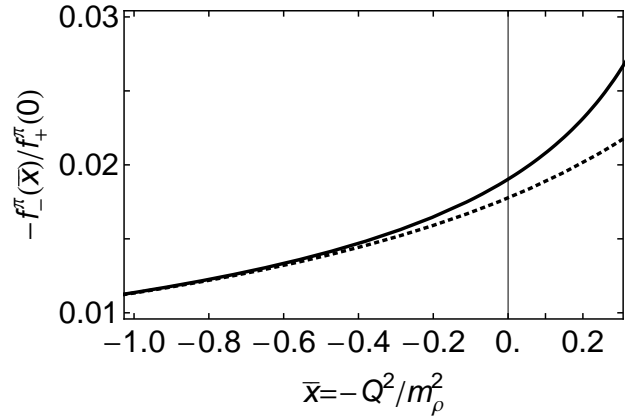}}
\caption{
$\pi_{e 3}$ transition form factor $f_-^\pi$.
Solid curve -- our contact interaction result; and dotted curve  -- $(1/5)f_-^K/f_+^K$ from Fig.\,\protect\ref{fig:FKl3m}.
\label{fig:Fpiem}}
\end{figure}

In Fig.\,\ref{fig:Fpiem} we plot $f_-^\pi$ and compare it with $(1/5)f_-^K/f_+^K$.  It is plain and understandable [see Eq.\,\eqref{answerScalar}] that $f_-^\pi$ exhibits a pole at the mass of a $u \bar u + d \bar d$-scalar meson, which is lighter than that at which the $u\bar s$-scalar pole appears in $f_-^K$.

We note that
\begin{equation}
f_-^\pi(0) = 0.019 \approx 2 \frac{M_d-M_u}{\Lambda_{\rm uv}} \approx 2 \frac{m_d-m_u}{\Lambda_{\rm uv}}\,.
\end{equation}
Comparison with Eq.\,\eqref{fminusK0} and $f_-^{B\to K}(0) = 0.28$, $f_-^{B\to \pi}(0)= 0.29$, from Ref.\,\cite{Ivanov:2007cw}, shows that whilst $f_-(0)$ is a gauge of flavour symmetry breaking, it is also sensitive to the difference between explicit and dynamically generated mass; viz., the rate at which $f_-(0)$ increases is damped as one moves away from the chiral limit and by the difference $M_f-m_f$.

Owing to their definition via the ratios in Eqs.\,\eqref{fpfmratios} and the kinematic constraints of the $\pi_{e 3}$ decay, in computing $f_\pm^\pi$ one must exercise particular care with numerical analysis, especially in calculating the contribution to $P\cdot M^{\pi_{e3}}$ from the $P_{1L}(Q^2)$-term in Eq.\,\eqref{genvector}.  In order to mitigate the difficulty when evaluating the $\pi_{e3}$ analogue of the integral in Eq.\,\eqref{ExplicitMKl3}, in the $P_{1L}^{du}(Q^2)$-term alone we employ isospin-averaged dressed-quark and pion masses.

\section{Epilogue}
\label{sec:epilogue}
Using the leading-order in a global-symmetry-preserving truncation of QCD's Dyson-Schwinger equations, we calculated elastic and semileptonic transition form factors for the kaon and pion. In these computations we employed a momentum-independent form for the leading-order kernel in the gap- and Bethe-Salpeter equations.  Amongst our results, we provide interpolations of the form factors.  They should be useful in working toward the broader aim of charting the interaction between light-quarks by explicating the impact of differing assumptions about the behaviour of the Bethe-Salpeter kernel upon the spectrum of hadrons, and also upon their elastic and transition form factors on a large domain of momentum transfer.

To provide a context for our results and to assist in understanding them, we compared our form factors with those obtained using the same truncation but an interaction that preserves the one-loop renormalisation group behaviour of QCD.  The comparison showed that in connection with experimental observables revealed by probes with $|Q^2|\lesssim M^2$, where $M\approx 0.4\,$GeV is an infrared value of the dressed-quark mass, results obtained using a symmetry-preserving regularisation of the contact-interaction are not realistically distinguishable from those produced by more sophisticated kernels.  It is notable, too, that available data on the kaon form factors do not extend into the domain whereupon one could distinguish between contact-interaction results and those obtained with QCD-like kernels.  These remarks may be quantified by the following observation: considering a collection of eleven $|Q^2|\lesssim M^2$ quantities, the contact interaction produces a rms-error relative to experiment of 25\%, whereas this error is $19$\% for the more sophisticated interaction.  (Expressed differently, the mean absolute value of the relative error (mmre) is $21\pm 10$\% cf.\ $11 \pm 11$\%.)

The picture is different if one includes the domain $Q^2>M^2$, whereupon a consistent treatment of the contact interaction yields harder form factors than those obtained with one-loop QCD renormalisation-group-improved kernels.  This owes to the necessary presence in pseudoscalar meson Bethe-Salpeter amplitudes of terms that may be described as pseudovector in character.

In this context, an inconsistent treatment of the contact interaction is possible; namely, through deliberate omission of the pseudoscalar mesons' pseudovector components.  Results obtained thereby are just as soft as those produced by a fully-consistent treatment of a momentum-dependent kernel that behaves as $1/k^2$ in the ultraviolet.

In the past, this omission was mere negligence; and results obtained were often misinterpreted as questioning the need for QCD.  From a modern perspective, however, the omission might be used judiciously in order to build efficacious models for hadron physics phenomena that cannot readily be studied using more elaborate means, so long as neither agreement nor disagreement with experiment is interpreted as a challenge to QCD.  For the phenomena studied herein, such an artifice is quite fruitful; e.g., it produces a rms-error of $15$\% (or mmre$\,=10\pm 9$\%) over the basket of eleven $|Q^2|\lesssim M^2$ quantities mentioned above and a $Q^2$-dependence of form factors that is typically almost indistinguishable from that obtained with the fully-consistent treatment of a sophisticated interaction.  In addition, it enables one to develop a perspective on the point at which perturbative-QCD behaviour might become apparent in meson form factors.

In stepping toward these conclusions, we were able to make a number of other observations.  For example, it was necessary to detail the properties of the inhomogeneous vector and scalar vertices, a process which led us to a novel Ward identity for the scalar vertex.  In addition, we found that the charge distribution of a dressed-$u$-quark in the $K^+$ is very similar to that of the dressed-$u$-quark in the $\pi^+$, whereas the charge distribution of the dressed-$s$-quark in the $K^+$ is noticeably harder than that of its $u$-quark partner.  This explains the positive slope of the $K^0$ form factor at $Q^2=0$.
Finally, whilst the $Q^2=0$ value of the subleading transition form factor, $f_-$, is a gauge of flavour symmetry breaking, it is also sensitive to the difference between the explicit current-quark mass and the interaction-generated dynamical mass.

This study lays a foundation for the contact-interaction computation of elastic and transition form factors involving baryons with strangeness.  It also emphasises that studies employing a symmetry-preserving regularisation of the contact interaction can usefully serve as a surrogate, enabling the exploration of domains which analyses using interactions that more closely follow the pointwise behaviour anticipated of QCD are not yet able to enter.  At present, prudent studies of this type are critical in attempts to use experimental data as a tool for charting the nature of the quark-quark interaction at long-range; i.e., for identifying distinct signals of the running of couplings and masses in QCD.

\section*{Acknowledgments}
%
We are grateful for valuable input from A.~Bashir, I.\,C.~Clo\"et, J.~Segovia~Gonz\'ales and P.\,C.~Tandy.
C.~Chen acknowledges the support of the China Scholarship Council (file no.\,2010634019).  This work was otherwise supported by:
Forschungszentrum J\"ulich GmbH;
and
U.\,S.\ Department of Energy, Office of Nuclear Physics, contract nos.~DE-AC02-06CH11357 and DE-SC0006765.

\appendix
\section{Contact interaction}
\label{sec:contact}
The key elements in our analysis are the dressed-quark propagators, the meson Bethe-Salpeter amplitudes and the quark--gauge-boson vertices.  All are completely determined once the quark-quark interaction kernel is specified.  We use
\begin{equation}
\label{njlgluon}
g^2 D_{\mu \nu}(p-q) = \delta_{\mu \nu} \frac{4 \pi \alpha_{\rm IR}}{m_G^2}\,,
\end{equation}
where $m_G=0.8\,$GeV is a gluon mass-scale typical of the one-loop renormalisation-group-improved interaction detailed in Ref.\,\cite{Qin:2011dd}, and the fitted parameter $\alpha_{\rm IR} = 0.93 \pi$ is commensurate with contemporary estimates of the zero-momentum value of a running-coupling in QCD \cite{Aguilar:2009nf,Oliveira:2010xc,Aguilar:2010gm,Boucaud:2010gr,Pennington:2011xs,Wilson:2012em}.  We embed Eq.\,\eqref{njlgluon} in a rainbow-ladder truncation of the DSEs.  This means
\begin{equation}
\label{RLvertex}
\Gamma_{\nu}(p,q) =\gamma_{\nu}
\end{equation}
in the gap equation and in the subsequent construction of the Bethe-Salpeter kernels.

The interaction in Eq.\,(\ref{njlgluon}) may be viewed as being inspired by models of the Nambu--Jona-Lasinio type \cite{Nambu:1961tp}.  Our treatment is atypical, however.  Used to build a rainbow-ladder truncation of the DSEs, Eqs.\,\eqref{njlgluon}, (\ref{RLvertex}) produce results for low-momentum-transfer observables that are directly comparable with those produced by more sophisticated interactions, as illustrated in Refs.\,\cite{Roberts:2011wy,Wilson:2011aa,GutierrezGuerrero:2010md,Roberts:2010rn,Chen:2012qr}.

\subsection{Gap equation}
\label{sec:gap}
The dressed-quark propagators in Eq.\,\eqref{ExplicitMKl3} are obtained from the gap equation:
\begin{equation}
 S_f^{-1}(p) =  i \gamma \cdot p + m_f +  \frac{16\pi}{3}\frac{\alpha_{\rm IR}}{m_G^2} \int\!\frac{d^4 q}{(2\pi)^4} \,
\gamma_{\mu} \, S_f(q) \, \gamma_{\mu}\,,   \label{gap-1}
\end{equation}
where $m_f$ is the quark's current-mass.  When the divergence is regularised in a Poincar\'e covariant manner, the solution is
\begin{equation}
\label{genS}
S_f(p)^{-1} = i \gamma\cdot p + M_f\,,
\end{equation}
where $M_f$ is momentum-independent and determined by
\begin{equation}
M_f = m_f + M_f\frac{4\alpha_{\rm IR}}{3\pi m_G^2} \int_0^\infty \!ds \, s\, \frac{1}{s+M_f^2}\,.
\end{equation}

\begin{table}[t]
\caption{\label{Tab:DressedQuarks}
Dressed-quark properties at $T=0$, computed from the gap equation and required as input for the Bethe-Salpeter equations.
All results obtained with $\alpha_{\rm IR} =0.93 \pi$ and (in GeV) $\Lambda_{\rm ir} = 0.24\,$, $\Lambda_{\rm uv}=0.905$.
(These parameters take the values determined in the spectrum calculation of Ref.\,\protect\cite{Roberts:2011cf}, which produces $m_\rho=0.928\,$GeV; isospin symmetry is assumed; and all dimensioned quantities are listed in GeV.)}
\begin{center}
\begin{tabular*}
{\hsize}
{
c@{\extracolsep{0ptplus1fil}}
c@{\extracolsep{0ptplus1fil}}
c@{\extracolsep{0ptplus1fil}}
c@{\extracolsep{0ptplus1fil}}
c@{\extracolsep{0ptplus1fil}}
c@{\extracolsep{0ptplus1fil}}
c@{\extracolsep{0ptplus1fil}}
c@{\extracolsep{0ptplus1fil}}}\hline
$m_u$ & $m_s$ & $m_s/m_u$ & $M_0$ &   $M_u$ & $M_s$ & $M_s/M_u$  \\\hline 
0.007  & 0.17 & 24.3 & 0.36 & 0.37 & 0.53 & 1.43  
\\\hline
\end{tabular*}
\end{center}
\end{table}

In regularising, we write \cite{Ebert:1996vx}
\begin{eqnarray}
\nonumber
\frac{1}{s+M^2} & = & \int_0^\infty d\tau\,{\rm e}^{-\tau (s+M^2)}  \\
&\rightarrow&  \int_{\tau_{\rm uv}^2}^{\tau_{\rm ir}^2} d\tau\,{\rm e}^{-\tau (s+M^2)}
\label{RegC}\\
 &=&
\frac{{\rm e}^{- (s+M^2)\tau_{\rm uv}^2}-{\rm e}^{-(s+M^2) \tau_{\rm ir}^2}}{s+M^2} \,, \label{ExplicitRS}
\end{eqnarray}
where $\tau_{\rm ir,uv}$ are, respectively, infrared and ultraviolet regulators.  It is apparent from the rightmost expression in Eq.\,(\ref{ExplicitRS}) that a finite value of $\tau_{\rm ir}=:1/\Lambda_{\rm ir}$ implements confinement by ensuring the absence of quark production thresholds \cite{Chang:2011vu,Bashir:2012fs}.  Since Eq.\,(\ref{njlgluon}) does not define a renormalisable theory, then $\Lambda_{\rm uv}:=1/\tau_{\rm uv}$ cannot be removed but instead plays a dynamical role, setting the scale of all dimensioned quantities.  Using Eq.\,\eqref{ExplicitRS}, the gap equation becomes
\begin{equation}
M_f = m_f + M_f\frac{4\alpha_{\rm IR}}{3\pi m_G^2}\,\,{\cal C}^{\rm iu}(M_f^2)\,,
\label{gapactual}
\end{equation}
where
\begin{eqnarray}
\nonumber
{\cal C}^{\rm iu}(\sigma) &= &
\int_0^\infty\! ds \, s \int_{\tau_{\rm uv}^2}^{\tau_{\rm ir}^2} d\tau\,{\rm e}^{-\tau (s+\sigma)}\\
& = &
\sigma \big[\Gamma(-1,\sigma \tau_{\rm uv}^2) - \Gamma(-1,\sigma \tau_{\rm ir}^2)\big],
\label{eq:C0}
\end{eqnarray}
with $\Gamma(\alpha,y)$ being the incomplete gamma-function.  It is convenient to define $\overline{\cal C}^{\rm iu}(\sigma)={\cal C}^{\rm iu}(\sigma)/\sigma$.

In Table~\ref{Tab:DressedQuarks} we report values of $u$- and $s$-quark properties, computed from Eq.\,\eqref{gapactual}, that will subsequently be used herein.  The input ratio $m_s/\bar m$, where $\bar m = (m_u+m_d)/2$, is consistent with contemporary estimates \cite{Leutwyler:2009jg}.  The result $M_s-m_s \approx M_0$ is typical \cite{Maris:1997tm,Bhagwat:2007ha} and indicates that the additive impact of DCSB is nearly as great for the $s$-quark as it is for $u,d$-quarks.  In general, however, $M_f-m_f$ is a monotonically decreasing function of $m_f$, bounded below by zero as $m_f\to\infty$ \cite{Bhagwat:2007ha,Holl:2005st}.
%

\subsection{Bethe-Salpeter equations}
In rainbow-ladder truncation and with the interaction in Eq.\,\eqref{njlgluon}, the homogeneous Bethe-Salpeter equation (BSE) for a meson comprised of quarks with flavours $f$, $\bar g$ is
\begin{equation}
\Gamma_{f\bar g}(Q) =  - \frac{16 \pi}{3} \frac{\alpha_{\rm IR}}{m_G^2}
\int \! \frac{d^4t}{(2\pi)^4} \gamma_\mu S_f(t+Q) \Gamma_{f\bar g}(Q)S_g(t) \gamma_\mu \,,
\label{LBSEI}
\end{equation}
where $Q$ is the total momentum of the bound-state.  This equation has a solution for $Q^2=-m_{f\bar g}^2$, where $m_{f\bar g}$ is the bound-state's mass.

The contact interaction supports a Bethe-Salpeter amplitude of the form
\begin{equation}
\label{KaonBSA}
\Gamma_{f\bar g}(P) = i \gamma_5 \,E_{f\bar g}(P) + \frac{1}{2 M_{f\bar g}} \gamma_5 \gamma\cdot P \, F_{f\bar g}(P)\,,
\end{equation}
where\footnote{The choice one makes for the mass-dimensioned constant $M_{f \bar g}$ has no effect on any result.}
$M_{f\bar g} = M_f M_{\bar g} /[M_f+M_{\bar g}]$.  If one inserts Eq.\,\eqref{KaonBSA} into Eq.\,\eqref{LBSEI} and employs the symmetry-preserving regularisation of the contact interaction explained, e.g., in Ref.\,\cite{Wilson:2011aa}, which requires
\begin{equation}
0 = \int_0^1d\alpha \,
\big[ {\cal C}^{\rm iu}(\omega_{f\bar g}(\alpha,Q^2))
%
+ \, {\cal C}^{\rm iu}_1(\omega_{f\bar g}(\alpha,Q^2))\big], \label{avwtiP}
\end{equation}
where ($\hat \alpha = 1-\alpha$)
\begin{eqnarray}
\omega_{f\bar g}(\alpha,Q^2) &=& M_f^2 \hat \alpha + \alpha M_{\bar g}^2 + \alpha \hat\alpha Q^2\,,
\label{eq:omega}
\end{eqnarray}
and
\begin{equation}
{\cal C}^{\rm iu}_1(z) = - z (d/dz){\cal C}^{\rm iu}(z)
= z\big[ \Gamma(0,M^2 \tau_{\rm uv}^2)-\Gamma(0,M^2 \tau_{\rm ir}^2)\big] ,\rule{2em}{0ex}
\label{eq:C1}
\end{equation}
then one arrives at the following explicit form of the Bethe-Salpeter equation:
\begin{equation}
\label{bsefinalE}
\left[
\begin{array}{c}
E_{f\bar g}(Q)\\
F_{f\bar g}(Q)
\end{array}
\right]
= \frac{4 \alpha_{\rm IR}}{3\pi m_G^2}
\left[
\begin{array}{cc}
{\cal K}_{EE}^{f\bar g} & {\cal K}_{EF}^{f\bar g} \\
{\cal K}_{FE}^{f\bar g} & {\cal K}_{FF}^{f\bar g}
\end{array}\right]
\left[\begin{array}{c}
E_{f\bar g}(Q)\\
F_{f\bar g}(Q)
\end{array}
\right],
\end{equation}
with
\begin{subequations}
\label{fgKernel}
\begin{eqnarray}
\nonumber
{\cal K}_{EE}^{f\bar g} &=&
\int_0^1d\alpha \bigg\{
{\cal C}^{\rm iu}(\omega_{f\bar g}( \alpha, Q^2))  \\
&&+ \bigg[ M_f M_{\bar g}-\alpha \hat\alpha Q^2 - \omega_{f\bar g}( \alpha, Q^2)\bigg]\nonumber \\
&&
\quad \times
\overline{\cal C}^{\rm iu}_1(\omega_{f\bar g}(\alpha, Q^2))\bigg\},\\
\nonumber
{\cal K}_{EF}^{f\bar g} &=& \frac{Q^2}{2 M_{f\bar g}} \int_0^1d\alpha\, \bigg[\hat \alpha M_f+\alpha M_{\bar g}\bigg]\\
&& \quad \times \overline{\cal C}^{\rm iu}_1(\omega_{f\bar g}(\alpha, Q^2)),\\
{\cal K}_{FE}^{f\bar g} &=& \frac{2 M_{f\bar g}^2}{Q^2} {\cal K}_{EF}^{f\bar g} ,\\
\nonumber
{\cal K}_{FF}^{f\bar g} &=& - \frac{1}{2} \int_0^1d\alpha\, \bigg[ M_f M_{\bar g}+\hat\alpha M_f^2+\alpha M_{\bar g}^2\bigg]\\
&& \quad \times \overline{\cal C}^{\rm iu}_1(\omega_{f\bar g}(\alpha, Q^2))\,.
\end{eqnarray}
\end{subequations}

Equation~\eqref{bsefinalE} is an eigenvalue problem, which has a solution for $Q^2=-m_{f\bar g}^2$, at which point the eigenvector is the meson's Bethe-Salpeter amplitude.  In the computation of observables one must employ the canonically normalised amplitude; viz., the amplitude rescaled such that
\begin{equation}
\label{normcan}
1=\left. \frac{d}{d Q^2}\Pi_{f \bar g}(Z,Q)\right|_{Z=Q},
\end{equation}
where
\begin{eqnarray}
\nonumber \Pi_{f\bar g}(Z,Q) &=& 6 {\rm tr}_{\rm D} \!\! \int\! \frac{d^4t}{(2\pi)^4} \Gamma_{f\bar g}(-Z)\\
&& \quad \times
 S_s(t+Q) \, \Gamma_{f\bar g}(Z)\, S_u(t)\,.
\end{eqnarray}

With the amplitudes and propagators in hand, one may compute all properties of the pions and kaons in rainbow-ladder truncation.  For example, the leptonic decay constants of the charged mesons and the in-meson condensates \cite{Maris:1997tm,Brodsky:2012ku} are respectively expressed:
\begin{eqnarray}
f_{f\bar g} &=& \frac{N_c}{4\pi^2}\frac{1}{ M_{f\bar g}}\,
\big[ E_{f\bar g} {\cal K}_{FE}^{f\bar g} + F_{f\bar g}{\cal K}_{FF}^{f\bar g} \big]\,, \label{ffg}\\
\kappa_{f\bar g} &=& f_{f\bar g} \frac{N_c}{4\pi^2} \big[E_{f\bar g} {\cal K}_{EE}^{f\bar g} + F_{f\bar g}{\cal K}_{EF}^{f\bar g} \big]\,, \label{kappafg}
\end{eqnarray}
wherein each quantity is computed at that $Q^2$ for which Eq.\,\eqref{bsefinalE} is satisfied for the meson under consideration.  In Table~\ref{Tab:BSE} we record computed values of all canonically normalised Bethe-Salpeter amplitudes relevant herein and results from Eqs.\,\eqref{ffg}, \eqref{kappafg}.
N.B.\, It is typical that the $\pi^+-\pi^0$ mass difference is small when electromagnetic contributions are neglected \cite{Bhagwat:2007ha,Gasser:1982ap}.

\begin{table}[t]
\caption{\label{Tab:BSE}
Selected meson properties, including the canonically normalised Bethe-Salpeter amplitudes, computed using the formulae described herein.  The row labelled ``chiral'' is obtained with $m_u=m_d=m_s=0$; and to compute that labelled $\pi^0$, we used $m_u=0.0029$, $m_d=0.011$, which produces $M_u=0.36$, $M_d=0.37$.  Note that $\bar m = (m_u+m_d)/2=0.007$, deliberately consistent with Table~\protect\ref{Tab:DressedQuarks}.
(All dimensioned quantities are listed in GeV.)}
\begin{center}
\begin{tabular*}
{\hsize}
{
c@{\extracolsep{0ptplus1fil}}
c@{\extracolsep{0ptplus1fil}}
c@{\extracolsep{0ptplus1fil}}
c@{\extracolsep{0ptplus1fil}}
c@{\extracolsep{0ptplus1fil}}
c@{\extracolsep{0ptplus1fil}}}\hline
    & $E_{f\bar g}$ & $F_{f\bar g}$ & $m_{f\bar g}$ & $f_{f\bar g}$ & $\kappa^{1/3}_{f\bar g}$\\\hline
\mbox{chiral} & 3.57 & 0.46 & 0 & 0.10 & 0.24 \\
$\pi^+$ & 3.64 & 0.48 & 0.14\phantom{9} & 0.10 & 0.24 \\
$\pi^0$ & 3.60 & 0.48 & 0.139 & & 0.24 \\
$K^+ $ & 3.82 & 0.59 & 0.50\phantom{9} & 0.11 & 0.25 \\\hline
\end{tabular*}
\end{center}
\end{table}

\section{Form-Factor Formulae}
\label{app:fff}
Equation\,\eqref{formulaFuKp} decomposes the $u$-quark contribution to the charged-kaon's form factor into a sum of three pieces, each associated with a different pairing of the terms in the kaon's Bethe-Salpeter amplitude:
\begin{subequations}
\label{eqs:TKEE}
{\allowdisplaybreaks
\begin{eqnarray}
\nonumber
T_{K,EE}^u &=& \frac{N_c}{4\pi^2}\bigg[\int_0^1 \! d\alpha\,
\overline {\cal C}_1^{\rm iu}(\omega_{uu})\\
\nonumber &&+2\int_0^1\! d\alpha d\beta\,\alpha\,
\bigg( 2 M_s [M_u - M_s] \\
&& + \alpha [(M_u-M_s)^2+m_K^2] \bigg)
\overline{\cal C}_2^{\rm iu}(\omega_{1u\bar s})\bigg],\quad\\
\nonumber
T_{K,EF}^u &=& \frac{N_c}{4\pi^2}\frac{1}{M_{u\bar s}^2}\bigg[\int_0^1 \! d\alpha d\beta \, \alpha\,\bigg(
{\cal A}_{uu\bar{s} }^{EF}\,\overline{\cal C}^{iu}_{1}  (\omega_{1u \bar s})\\
&&
+[{\cal B}_{uu \bar{s}}^{EF} - {\cal A}_{uu \bar{s}}^{EF}\,
\omega_{1 u \bar s}]
\overline{\cal C}^{\rm iu}_{2}(\omega_{1 u \bar s})\bigg)\bigg],\\
\nonumber
T_{K,FF}^u &=& \frac{N_c}{4\pi^2}\frac{1}{M_{u\bar s}^2}\bigg[\int_0^1 \! d\alpha d\beta \, \alpha\,\bigg(
{\cal A}_{uu\bar{s} }^{FF}\,\overline{\cal C}^{iu}_{1}  (\omega_{1u \bar s})\\
&&
+[\frac{1}{2}{\cal B}_{uu \bar{s}}^{FF} - {\cal A}_{uu \bar{s}}^{FF}\,
\omega_{1 u \bar s}]
\overline{\cal C}^{\rm iu}_{2}(\omega_{1 u \bar s})\bigg)\bigg],
\end{eqnarray}}
\end{subequations}
\hspace*{-0.6\parindent}%
where: $\omega_{uu}=\omega_{uu}(\alpha,Q^2)$, with the latter defined in Eq.\,\eqref{eq:omega}; ($\hat\beta = 1-\beta$)
\begin{eqnarray}
\omega_{1u\bar s} &=& \alpha M_u^2 + \hat\alpha M_{s}^2 + \alpha^2\beta\hat\beta \, Q^2 - \alpha\hat\alpha\, m_K^2\,,\\
{\cal C}_2^{\rm iu}(\sigma) &=& \frac{\sigma^2}{2}  \frac{d^2}{d\sigma^2} {\cal C}^{\rm ir}(\sigma) = \frac{\sigma}{2}\big[ {\rm e}^{-\sigma r_{\rm uv}^2} - {\rm e}^{-\sigma r_{\rm ir}^2}\big]\,,
\end{eqnarray}
with $\overline{\cal C}_2^{\rm iu}(\sigma)={\cal C}_2^{\rm iu}(\sigma)/\sigma^2$; and
\begin{subequations}
\begin{eqnarray}
{\cal A}_{uu\bar{s} }^{EF} &=& -4 M_{u\bar s} [\theta M_u - (1/2-\theta) M_s]\,,
\label{AEFtheta}\\
{\cal A}_{uu\bar{s} }^{FF} &=& [1/2-\theta][(\alpha-2\hat\alpha)m_K^2+\alpha Q^2]\,,
\label{AFFtheta}\\
\nonumber
{\cal B}_{uu\bar{s} }^{EF} &=&
-M_{u\bar s}(
2 [M_s M_u^2 + m_K^2 \hat\alpha( M_s \hat\alpha + 2\alpha M_u)]\\
&&-\alpha[\alpha M_u + M_s (\hat\alpha + 2\alpha\beta\hat\beta)]Q^2)\,,\\
\nonumber
{\cal B}_{uu\bar{s} }^{EF} &=&
m_K^2 [ 2 M_s M_u\hat\alpha + \alpha(M_u^2+m_K^2\hat\alpha^2)]\\
\nonumber && - \alpha Q^2 [ M_u M_s + \alpha m_K^2 (\hat\alpha + \beta\hat\beta [\alpha-2\hat\alpha]) ]\,.\\
&&
\end{eqnarray}
\end{subequations}
The parameter $\theta$ is explained in App.\,\ref{app:current}.

\section{Current conservation}
\label{app:current}
In deriving the formulae in Sec.\,\ref{app:fff} we followed the methods detailed in Refs.\,\cite{GutierrezGuerrero:2010md,Roberts:2011wy} and indicated in Sec.\,\ref{sec:WVV} herein. They rely in part on O(4) invariance and the assumption of a translationally invariant regularisation of the integrals.  The latter is always formally true but, in the presence of pseudovector components in the pseudoscalar meson, it is practically broken with the contact interaction once Eq.\,\eqref{RegC} is used.  The effect is to produce a small nonzero result for $[T^u_{K,EF}(Q^2=0)-T^s_{K,EF}(Q^2=0)]$, typically a relative error of $\lesssim 1\,$\%, whereas this difference should always vanish.

The weakness can be traced to the quadratic divergences that arise through integrals such as
\begin{eqnarray}
\nonumber
&&\int\frac{d^4 t}{(2\pi)^4} \frac{1}{[t^2+\omega]^2} \{(P\cdot t)^2,(Q\cdot t)^2, (P\cdot t) (Q\cdot t)\}\\
&=& \int\frac{d^4 t}{(2\pi)^4} \frac{t^2}{[t^2+\omega]^2}\frac{1}{4}
\{P^2,Q^2 , P\cdot Q \}\,, \label{eq:1on4}
\end{eqnarray}
which also affect the value of $f_+^K(0)$.  It can be ameliorated via a simple expedient: in Eq.\,\eqref{eq:1on4}, replace
\begin{equation}
1/4 \to \theta = 1.456 (1/4)\,.
\label{eq:theta}
\end{equation}
This is the origin and value of $\theta$ in Eqs.\,\eqref{AEFtheta}, \eqref{AFFtheta}.  We employ the same improvisation in connection with the $K_{\ell 3}$ form factors.


\end{document}